\documentclass[ALICE,manyauthors]{cernphprep}
\usepackage[comma,square,numbers,sort&compress]{natbib}

\usepackage{lineno}
\usepackage{xspace}
\usepackage{hyperref}
\usepackage{bigdelim}

\usepackage[inline]{enumitem}
\usepackage[T1]{fontenc}
\usepackage{orcidlink}

\begin{document}
%%%%%%%%%%%%%%%%%%%%%%%%%%%%%%%%%%%%%%%%%%%%%%%%%%
% These are some new commands that may be useful 
% for paper writing in general. If other newcommands
% are needed for your specific paper, please feel 
% free to add here. 
%
% The currently available commands are organized in: 
% 1) Systems
% 2) Quantities
% 3) Energies and units
% 4) Detectors
% 5) particle species 
%%%%%%%%%%%%%%%%%%%%%%%%%%%%%%%%%%%%%%%%%%%%%%%%%%

% 1) SYSTEMS 
\newcommand{\pp}           {pp\xspace}
\newcommand{\ppbar}        {\mbox{$\mathrm {p\overline{p}}$}\xspace}
\newcommand{\XeXe}         {\mbox{Xe--Xe}\xspace}
\newcommand{\PbPb}         {\mbox{Pb--Pb}\xspace}
\newcommand{\pA}           {\mbox{pA}\xspace}
\newcommand{\pPb}          {\mbox{p--Pb}\xspace}
\newcommand{\Pbp}          {\mbox{Pb--p}\xspace}
\newcommand{\AuAu}         {\mbox{Au--Au}\xspace}
\newcommand{\dAu}          {\mbox{d--Au}\xspace}

% 2) QUANTITIES 
\newcommand{\s}            {\ensuremath{\sqrt{s}}\xspace}
\newcommand{\snn}          {\ensuremath{\sqrt{s_{\mathrm{NN}}}}\xspace}
\newcommand{\pt}           {\ensuremath{p_{\rm T}}\xspace}
\newcommand{\meanpt}       {$\langle p_{\mathrm{T}}\rangle$\xspace}
\newcommand{\ycms}         {\ensuremath{y_{\rm cms}}\xspace}
\newcommand{\ylab}         {\ensuremath{y_{\rm lab}}\xspace}
\newcommand{\etacms}       {\ensuremath{\eta_{\rm cms}}\xspace}
\newcommand{\etalab}       {\ensuremath{\eta_{\rm lab}}\xspace}
\newcommand{\etarange}[1]  {\mbox{$\left | \eta \right |~<~#1$}}
\newcommand{\yrange}[1]    {\mbox{$\left | y \right |~<~#1$}}
\newcommand{\dndy}         {\ensuremath{\mathrm{d}N_\mathrm{ch}/\mathrm{d}y}\xspace}
\newcommand{\dndeta}       {\ensuremath{\mathrm{d}N_\mathrm{ch}/\mathrm{d}\eta}\xspace}
\newcommand{\avdndeta}     {\ensuremath{\langle\dndeta\rangle}\xspace}
\newcommand{\dNdy}         {\ensuremath{\mathrm{d}N_\mathrm{ch}/\mathrm{d}y}\xspace}
\newcommand{\Npart}        {\ensuremath{N_\mathrm{part}}\xspace}
\newcommand{\Ncoll}        {\ensuremath{N_\mathrm{coll}}\xspace}
\newcommand{\dEdx}         {\ensuremath{\textrm{d}E/\textrm{d}x}\xspace}
\newcommand{\RpPb}         {\ensuremath{R_{\rm pPb}}\xspace}

% 3) ENERGIES, UNITS
\newcommand{\nineH}        {$\sqrt{s}~=~0.9$~Te\kern-.1emV\xspace}
\newcommand{\thirteen}        {$\sqrt{s}~=~13$~Te\kern-.1emV\xspace}
\newcommand{\seven}        {$\sqrt{s}~=~7$~Te\kern-.1emV\xspace}
\newcommand{\twoH}         {$\sqrt{s}~=~0.2$~Te\kern-.1emV\xspace}
\newcommand{\twosevensix}  {$\sqrt{s}~=~2.76$~Te\kern-.1emV\xspace}
\newcommand{\five}         {$\sqrt{s}~=~5.02$~Te\kern-.1emV\xspace}
\newcommand{\twosevensixnn}{$\sqrt{s_{\mathrm{NN}}}~=~2.76$~Te\kern-.1emV\xspace}
\newcommand{\fivenn}       {$\sqrt{s_{\mathrm{NN}}}~=~5.02$~Te\kern-.1emV\xspace}
\newcommand{\LT}           {L{\'e}vy-Tsallis\xspace}
\newcommand{\GeVc}         {Ge\kern-.1emV/$c$\xspace}
\newcommand{\MeVc}         {Me\kern-.1emV/$c$\xspace}
\newcommand{\TeV}          {Te\kern-.1emV\xspace}
\newcommand{\GeV}          {Ge\kern-.1emV\xspace}
\newcommand{\MeV}          {Me\kern-.1emV\xspace}
\newcommand{\GeVmass}      {Ge\kern-.2emV/$c^2$\xspace}
\newcommand{\MeVmass}      {Me\kern-.2emV/$c^2$\xspace}
\newcommand{\lumi}         {\ensuremath{\mathcal{L}}\xspace}

% 4) DETECTORS 
\newcommand{\ITS}          {\rm{ITS}\xspace}
\newcommand{\TOF}          {\rm{TOF}\xspace}
\newcommand{\ZDC}          {\rm{ZDC}\xspace}
\newcommand{\ZDCs}         {\rm{ZDCs}\xspace}
\newcommand{\ZNA}          {\rm{ZNA}\xspace}
\newcommand{\ZNC}          {\rm{ZNC}\xspace}
\newcommand{\SPD}          {\rm{SPD}\xspace}
\newcommand{\SDD}          {\rm{SDD}\xspace}
\newcommand{\SSD}          {\rm{SSD}\xspace}
\newcommand{\TPC}          {\rm{TPC}\xspace}
\newcommand{\TRD}          {\rm{TRD}\xspace}
\newcommand{\VZERO}        {\rm{V0}\xspace}
\newcommand{\TZERO}        {\rm{T0}\xspace}
\newcommand{\VZEROA}       {\rm{V0A}\xspace}
\newcommand{\VZEROC}       {\rm{V0C}\xspace}
\newcommand{\Vdecay} 	   {\ensuremath{V^{0}}\xspace}

% 4) PARTICLE SPECIES 
\newcommand{\ee}           {\ensuremath{e^{+}e^{-}}} 
\newcommand{\pip}          {\ensuremath{\pi^{+}}\xspace}
\newcommand{\pim}          {\ensuremath{\pi^{-}}\xspace}
\newcommand{\kap}          {\ensuremath{\rm{K}^{+}}\xspace}
\newcommand{\kam}          {\ensuremath{\rm{K}^{-}}\xspace}
\newcommand{\pbar}         {\ensuremath{\rm\overline{p}}\xspace}
\newcommand{\kzero}        {\ensuremath{{\rm K}^{0}_{\rm{S}}}\xspace}
\newcommand{\lmb}          {\ensuremath{\Lambda}\xspace}
\newcommand{\almb}         {\ensuremath{\overline{\Lambda}}\xspace}
\newcommand{\Om}           {\ensuremath{\Omega^-}\xspace}
\newcommand{\Mo}           {\ensuremath{\overline{\Omega}^+}\xspace}
\newcommand{\X}            {\ensuremath{\Xi^-}\xspace}
\newcommand{\Ix}           {\ensuremath{\overline{\Xi}^+}\xspace}
\newcommand{\Xis}          {\ensuremath{\Xi^{\pm}}\xspace}
\newcommand{\Oms}          {\ensuremath{\Omega^{\pm}}\xspace}
\newcommand{\degree}       {\ensuremath{^{\rm o}}\xspace}

% OWN COMMANDS
\newcommand{\Jpsi}{\rm{J}/\psi}
\newcommand{\psit}{\psi(2{\rm S})}
\newcommand{\psip}{\psi(2{\rm S})}
\newcommand{\psipOJpsi}{\ensuremath{\psip\mbox{-over-}\Jpsi}}
\newcommand{\abs}[1]{\left\vert #1 \right\vert}
\newcommand{\ave}[1]{\langle #1 \rangle}
\newcommand{\zv}{z_{\rm vtx}}
\newcommand{\Ntr}{N_{\rm tracklet}}
\newcommand{\avNtr}{\langle N_{\rm tracklet} \rangle}
\newcommand{\Ntrc}{N_{\rm tracklet}^{\rm corr}}
\newcommand{\Ncorr}        {\ensuremath{N^\mathrm{corr}}\xspace}
\newcommand{\Ae}{A\varepsilon}
\newcommand{\mumu}{\mu^+\mu^-}
\newcommand{\BR}{\rm{BR}_{\Jpsi\rightarrow\mumu}}

\newcommand{\Ups}[1]{\Upsilon(#1{\rm S})}
\newcommand{\eightnn}       {$\sqrt{s_{\mathrm{NN}}}~=~8.16$~Te\kern-.1emV\xspace}
\newcommand{\AAcoll}          {\mbox{A--A}\xspace}

%%%%%%%%%%%%%%%  Title page %%%%%%%%%%%%%%%%%%%%%%%%
\begin{titlepage}
% the dates below correspond to CERN approval
% please don't touch: EB chairs will take care
\PHyear{2022}       % required, will be obtained from CERN
\PHnumber{064}      % required, will be obtained from CERN
\PHdate{24 March}  % required, will be obtained from CERN
%%%%%%%%%%%%%%%%%%%%%%%%%%%%%%%%%%%%%%%%%%%%%%%%%%%%

%%% Put your own title + short title here:
\title{Measurement of $\psip$ production as a function of charged-particle pseudorapidity density in pp collisions at $\s = 13$ TeV and p--Pb collisions at $\snn = 8.16$ TeV with ALICE at the LHC}
\ShortTitle{$\psip$ production vs. charged-particle multiplicity in pp and p--Pb collisions}   % appears on left page headers

%%% Do not change the next lines
\Collaboration{ALICE Collaboration\thanks{See Appendix~\ref{app:collab} for the list of collaboration members}}
\ShortAuthor{ALICE Collaboration} % appears on right page headers, do not change

\begin{abstract}
Production of inclusive charmonia in pp collisions at center-of-mass energy of $\s = 13$ TeV and p--Pb collisions at center-of-mass energy per nucleon pair of $\snn = 8.16$ TeV
is studied as a function of charged-particle pseudorapidity density with ALICE. 
Ground and excited charmonium states ($\Jpsi$, $\psip$) are measured from their dimuon decays in the interval of rapidity in the center-of-mass frame $2.5<\ycms<4.0$ for pp collisions, and $2.03<\ycms<3.53$ and $-4.46<\ycms<-2.96$ for p--Pb collisions. The charged-particle pseudorapidity density is measured around midrapidity ($|\eta|<1.0$). 
In \pp collisions, the measured charged-particle multiplicity extends to about six times the average value, while in \pPb collisions at forward (backward) rapidity a multiplicity corresponding to about three (four) times the average is reached.
The $\psip$  yield increases with the  charged-particle pseudorapidity density. 
The  ratio of $\psip$ over $\Jpsi$ yield does not show a significant multiplicity dependence in
either colliding system, suggesting a similar behavior of $\Jpsi$ and $\psip$ yields with respect to charged-particle pseudorapidity density. 
Results for the $\psip$ yield and its ratio with respect to $\Jpsi$ agree with available model calculations.
\end{abstract}
\end{titlepage}

\setcounter{page}{2} %please do not remove this line

%%%%%%%%%%%%%%%%%%%%%%%%%%%%%%%%
% begin main text
%%%%%%%%%%%%%%%%%%%%%%%%%%%%%%%%

% \newpage

\section{Introduction}
\label{sec:Intro}

Quarkonium production in hadronic collisions is a complex mechanism involving hard-scale processes, i.e., the creation of the quark--antiquark pair in the initial hard scattering, as well as the subsequent soft-scale process of the binding of the pair into a colorless final state~\cite{Bodwin:1994jh}. 
The production mechanism is sensitive to the gluon content of the colliding hadrons and thus to the parton distributions of the incoming proton (PDF) or nucleus (nPDF)~\cite{Ma:2014mri,PhysRevC.71.054902}. 
In collisions involving heavy nuclei, modification to the production with respect to that in \pp collisions may also arise from, e.g., energy loss that the heavy quarks experience while traversing the nucleus~\cite{Arleo:2014oha} or from subsequent interactions of the final states with comoving matter~\cite{Ferreiro:2014bia}, both of which are expected to lead to a suppression of the quarkonium yield. 
Good understanding of these phenomena is imperative in order to correctly interpret the data from nucleus--nucleus collisions, where quarkonia are expected to be suppressed due to a deconfined partonic medium, i.e., the quark--gluon plasma (QGP)~\cite{Matsui:1986dk}. At the energies reached by the Large Hadron Collider (LHC), the suppression is partially compensated by a regeneration of the bound states in the medium~\cite{ALICE:2019nrq}. 
% %

Measurements of quarkonium production at the LHC in small collision systems, i.e., proton--proton (\pp) and proton--nucleus (\pPb) collisions, shed more light onto these processes. While measurements in \pp  collisions allow one to study the baseline production mechanisms of the quark--antiquark pair, the minimum-bias \pPb data serve to probe the nuclear effects in conditions at which a formation of an extended QGP phase is not expected. 
In nuclear collisions, the initial-state nuclear effects impact the quark--antiquark pair created in the hard scattering. These effects manifest in the form of an enhancement and/or suppression of quarkonium production with respect to that in collisions of protons. Various initial-state effects are able to reproduce the measured nuclear modification of the  $\Jpsi$ and $\Ups{1}$ yields in \pPb collisions at the LHC (see Refs.~\cite{ALICE:2018mml,ALICE:2018szk,ALICE:2019qie,LHCb:2017ygo,LHCb:2018psc,CMS:2017exb} and references therein).
On the other hand, the excited charmonia ($\psip$) and bottomonia ($\Ups{2}$, $\Ups{3}$) display a suppression pattern different from that of their respective tighter bound states $\Jpsi$ and $\Ups{1}$~\cite{ALICE:2020vjy,LHCb:2018psc,LHCb:2016vqr}. Namely, the excited states were found to be more suppressed than the ground states.
Such behavior cannot be explained with initial-state effects only and suggests that additional, final-state effects, which act on a bound quarkonium state, need to be considered. Furthermore, the nuclear effects have been found to depend on the multiplicity of particles produced in the \pPb collision~\cite{ALICE:2015kgk,ALICE:2016sdt,ALICE:2017wet,ALICE:2020tsj}. ALICE Collaboration has previously published a measurement, complementary to the one presented in hereby work, of $\Jpsi$ and $\psip$ production as a function of centrality in \pPb collisions at \eightnn ~\cite{ALICE:2020tsj}. The study revealed that, except for the events in the lowest and highest event activity class, in which the nuclear modification between the two states is compatible, the $\psip$ yield is more suppressed than the $\Jpsi$ one. This result is consistent with the picture of the excited $\psip$ being more sensitive to final-state interactions. 

It is also interesting to note that the $\psip$ is rather unique among other quarkonia as it has negligible contribution to its production from decays of heavier states~\cite{ParticleDataGroup:2020ssz}. These so called feed-down contributions are accounted for in calculations used to model the suppression of $\Jpsi$ in the nuclear medium, namely contributions from $\chi_c(1{\rm P})$ and $\psip$. However, these models are impacted by the scarcity of available data on production of the excited charmonia in nuclear collisions, or lack thereof in the case of the P-wave states.

%%% multiplicity studies
Previous measurements of $\Jpsi$ production at forward rapidity as a function of multiplicity have revealed that the normalized yields increase linearly with the normalized charged-particle pseudorapidity density at midrapidity in \pp collisions~\cite{ALICE:2012pet,ALICE:2021zkd} (both quantities being normalized to their corresponding averages in minimum-bias events),
while for $\Jpsi$ production at midrapidity, the increase of the normalized yields has been found to be stronger than linear~\cite{ALICE:2020msa}.
In \pPb collisions, the trend also depends on the $\Jpsi$ rapidity and hints at a deviation from a linear trend, suggesting  slower- and stronger-than-linear increase for forward- and backward-rapidity $\Jpsi$, respectively~\cite{ALICE:2017wet,ALICE:2020eji}.
The measured multiplicity-dependent behavior of hidden-charm hadrons is compatible with that of open-charm and beauty hadrons, suggesting a common origin of these phenomena independent of hadronization~\cite{ALICE:2015ikl,ALICE:2020hdw}.

Moreover, measurements in high-multiplicity \pp and \pPb collisions have revealed the presence of phenomena typically attributed to creation of QGP, e.g., long-range near- and away-side ridges in two-particle azimuthal correlations~\cite{CMS:2010ifv,ALICE:2012eyl,ATLAS:2012cix,LHCb:2015coe}, collective motion of charged particles~\cite{CMS:2016fnw} and charmed hadrons~\cite{ALICE:2017smo,CMS:2018loe,CMS:2018duw}. 
Therefore, multiplicity-dependent studies in small systems provide a testing ground for examining the onset of QGP-like effects in collisions of energetic hadrons.

In this Letter, measurements of inclusive $\psip$ production at forward rapidity as a function of charged-particle pseudorapidity density at midrapidity in \pp and \pPb collisions are presented. The inclusive yield contains a \textit{prompt} component, which consists of states produced directly via hadronization of the quark--antiquark pair, and the \textit{non-prompt} component originating from decays of $b$-hadrons. The $\psip$ is compared with the $\Jpsi$ state, by measuring the ratio of $\psip$ to $\Jpsi$ yields as a function of multiplicity.
The data samples used in this study were collected with the ALICE detector at center-of-mass energies per nucleon pair of \thirteen for \pp and \eightnn for \pPb collisions.

\section{Detector and data samples}
\label{sec:Data}

The ALICE apparatus design and performance are documented in Refs.~\cite{Collaboration_2008,ALICE:2014sbx}. The following section only describes those detector subsystems which are relevant for the present analyses. 

The ALICE detector is instrumented at both central and forward rapidity. The ALICE forward muon spectrometer~\cite{ALICE:2013osk} detects muons in the pseudorapidity range \mbox{$-4<~\eta~<-2.5$} in the laboratory frame. It consists of five tracking stations (each of them consisting of two chambers of Cathode Pad Chamber detectors), followed by two triggering stations (two planes of Resistive Plate Chamber detectors each). A ten-interaction-length-thick absorber, made of carbon, concrete, and steel, is positioned in front of the tracking system to filter out most of the hadrons produced in the collisions. Remaining hadrons and low-momentum muons are absorbed by a second iron absorber positioned between the muon tracking chambers (MCH) and the muon trigger chambers (MTR). A 3 T·m dipole magnet, surrounding the third tracking station, provides the track bending for momentum evaluation. The Silicon Pixel Detector (\SPD) is part of the ALICE central barrel~\cite{ALICE:2010tia}. It is used to reconstruct the primary vertex, reject events with collision pile-up, and estimate the charged-particle multiplicity of the collision. It corresponds to the two innermost layers of the Inner Tracking System (\ITS)~\cite{ALICE:2010tia}, which are positioned around the beam pipe and cover the pseudorapidity intervals \etarange2 and \etarange1.4, respectively. 

The Minimum Bias (MB) trigger is provided by the \VZERO detector~\cite{ALICE:2013axi}, two scintillator arrays covering the pseudorapidity ranges \mbox{$2.8<~\eta~<5.1$} and \mbox{$-3.7<~\eta~<-1.7$}. The timing information from the \VZERO 
is used to remove beam-induced background. 
Finally, the luminosity determination is obtained from the \VZERO information and, independently, using the \TZERO Cherenkov detectors~\cite{TZEROdet}, which cover \mbox{$4.6 <~\eta~<4.9$} and \mbox{$-3.3 <~\eta~<-3.0$}~\cite{ALICE:2016nst,ALICE:2018ldk}.

The \pp data sample was collected at the center-of-mass energy of \thirteen between 2016 and 2018. Concerning the \pPb sample, the hereby presented data were collected in 2016 at the center-of-mass energy per nucleon pair of \eightnn. 
In these asymmetric collisions, the energy of the proton (Pb) beams is of $E_{\rm N}= 6.5$~(2.56) TeV per nucleon, and 
the center of mass of the nucleon--nucleon system is positioned at \mbox{$\Delta y= - 0.465$} in the laboratory frame.  
The \pPb data samples were recorded in two beam configurations: the \textit{forward} configuration, in which the proton moves toward the spectrometer and quarkonia are measured in the proton-going direction, i.e.,\ $2.03 < \ycms < 3.53$, and the \textit{backward} configuration, in which the proton moves away from the spectrometer and quarkonia are measured in the lead-going direction, i.e.,\ $-4.46 < \ycms < -2.96$. 
Events selected for these analyses were collected using a dimuon trigger which requires that two muons of opposite-sign charge are detected in the MTR in coincidence with the MB trigger, i.e., the detection of a signal in each side of the \VZERO. The muons were required to have a transverse momentum $\pt^{\rm trig}$, evaluated with the MTR, larger than about 0.5~\GeVc.
In these data-taking periods, the maximum pile-up probability  was about 0.5\% for \pp collisions, and about 4\% for \pPb collisions. Pile-up was further reduced to a negligible contribution by a dedicated event-selection strategy. An algorithm to tag events with multiple vertices reconstructed with the SPD was used for both collision systems. 
Additional selection criteria have been considered for \pPb data, i.e., the correlation between the number of clusters in SPD and the number of track segments in SPD, the correlation between the signal amplitude in the V0 and the number of track segments in SPD, a timing criterion on the signal from V0, and the correlation of timing signals from the two sides of the ZDC system. 
The integrated luminosity of the collected samples amounts to $24.38 \pm 0.87$~pb$^{-1}$ in \pp collisions and to $7.2 \pm 0.2$~nb$^{-1}$ ($10.6 \pm 0.3$~nb$^{-1}$) in forward (backward) \pPb collisions.

\section{Charged-particle multiplicity measurement}
\label{sec:multiplicity}

The number of track segments, \textit{tracklets}, ($\Ntr$) formed by combining hits in both SPD~\cite{ALICE:2010tia} layers pointing to the primary vertex was used as the estimator of the charged-particle pseudorapidity density (\dndeta) at midrapidity. The information provided by the SPD was also used to compute the position of the primary vertex. In order to reduce the potential effects of the detector non-uniformities at its acceptance limits, the position of the primary vertex along the beam axis ($z$) was restricted to \mbox{$|z_{\rm vtx}| < 10$ cm} and only tracklets within the pseudorapidity range \etarange1 were considered. 

To account for the limited acceptance of the SPD and the variation of its conditions over time, a data-driven event-by-event correction~\cite{ALICE:2012pet,ALICE:2015ikl,ALICE:2020eji,ALICE:2020msa} was applied to the raw number of tracklets. 
The dependence of the SPD acceptance on the vertex position was corrected for by dividing the raw number of tracklets in each event by a $z_{\rm vtx}$-dependent renormalization factor, which was defined for each data-taking period as the average value of the $\Ntr$ distribution in the corresponding $z_{\rm vtx}$ interval, $\avNtr$($z_{\rm vtx}$), normalized to a reference value. The reference value was chosen as the maximum value of $\avNtr$ over all $z_{\rm vtx}$ intervals and all data-taking periods.
In order to account for event-by-event fluctuations, the renormalization factor was randomly smeared for each event using a Poisson distribution. Given the variations of the SPD conditions with time, the dataset was split into groups of sub-periods with a similar SPD status. In particular, for the three-year-long  \pp data-taking period, 12 groups were considered. The  $\avNtr$($z_{\rm vtx}$) distributions were separately renormalized to the same reference value in each group. Once the correction was applied, events from all groups were merged and sorted into 9 intervals of corrected number of tracklets ($\Ntrc$). Considering the smaller \pPb data samples, 6 (8) $\Ntrc$ intervals were defined for the forward (backward) \pPb configuration in view of the $\psip$ signal extraction. 

The estimation of \dndeta from $\Ntrc$ was performed using Monte Carlo  (MC) simulations. DPMJET~\cite{Roesler:2000he} and PYTHIA~8.2~\cite{Sjostrand:2014zea} event generators were used to generate \pPb and \pp events respectively. In both cases, the  transport of the generated particles through the detector was simulated using GEANT3~\cite{Brun:1994aa}. The correlation between the generated \dndeta and the reconstructed $\Ntrc$ was parameterized with a second order polynomial function. Other types of functional form were also tested to take into account potential deviations from the assumed second order polynomial function. 

In the analysis of \pPb collisions, these deviations from the assumed quadratic parameterization were taken as one source of systematic uncertainty, ranging from  0.3$\%$ at intermediate multiplicity to 4$\%$ in the lowest multiplicity interval. Another considered source of systematic uncertainty related to the MC sample was the uncertainty on the residual $z_{\rm vtx}$ dependence of $\Ntrc$ originating from the differences between data and simulations, and the value of uncertainty was found to be close to 3$\%$. In addition, events generated with EPOS LHC~\cite{Pierog:2013ria} were also analyzed to evaluate the generator influence on the \dndeta determination, resulting in a 2$\%$ systematic uncertainty in all multiplicity intervals. The process to determine the two latter sources of uncertainty is identical to the work described in Ref.~\cite{ALICE:2020eji}.

In the case of \pp collisions, all three potential contributions to systematic uncertainty, described in the previous paragraph, were evaluated together as a single contribution. As in the \pPb case, EPOS LHC was considered to account for possible dependence on the choice of the event generator. The overall systematic uncertainty on \dndeta computed in each $\Ntrc$ interval was found to range between 0.9$\%$ (at intermediate multiplicity) and 5.3$\%$ (lowest multiplicity interval).

        \begin{table}[!tbhp]
        \centering
        \caption{Summary of systematic uncertainties on the charged-particle pseudorapidity density measurements in \etarange1 for \pp collisions at \thirteen and \pPb collisions at \eightnn. 
        In the case of \pp collisions, the contribution from the first three listed uncertainty sources was determined together as a single contribution, and listed in the middle. 
        Whenever a range of values is quoted, it refers to the variation as a function of multiplicity.
        }
        \begin{tabular}{l|c|c}
        \hline 
          & \pp & \pPb\\
        \hline
        vertex equalisation & \rdelim\}{3}{17.5mm}[\parbox{12.5mm}{0.9--5.3\%}] & 3\% \\
        \dndeta from $\Ntrc$ & & 0.3--4.0\% \\
        MC influence & & 2\% \\ \hline
        $\avdndeta$   & 1.4\%   & 4.1\% \\
        \hline 
        $\dndeta \left/\avdndeta \right.$ & 1.7--5.5\% & 5.4--6.5\%\\ \hline
        \end{tabular}
        \label{tab:sys_Nch}
    \end{table}

Finally, the average charged-particle pseudorapidity density in the non-single diffractive (NSD) \pPb collisions, $\avdndeta_{\rm NSD}$, was evaluated in an independent analysis. The NSD event class includes non- and double-diffractive events. The measurement, whose value in a narrower pseudorapidity range can be found in Ref.~\cite{ALICE:2018wma}, gives $\avdndeta_{\rm NSD}$ = 20.33 $\pm$ 0.83 (20.32 $\pm$ 0.83) for the forward (backward) configuration. 
The corresponding value for \pp collisions was measured for inelastic events with at least one charged particle at midrapidity (this class of events being commonly denoted as INEL$>$0) and amounts to $\avdndeta_{\rm INEL>0} = 7.07 ^{+0.10}_{-0.08}$,  computed from Ref.~\cite{ALICE:2020swj}. The total systematic uncertainties for both \pp and \pPb multiplicity measurements are reported in Table~\ref{tab:sys_Nch}.

\section{Charmonium yield determination}
\label{sec:yield}

The normalized charmoniun yields were defined as the corrected yields in a given charged-particle pseudorapidity density interval $i$, ${\rm d} N^i / {\rm d} y$, divided by their multiplicity-integrated values , $\langle {\rm d} N / {\rm d} y \rangle$, according to
    \begin{equation}
        \frac{ {\rm d} N^i / {\rm d} y } { \langle {\rm d} N / {\rm d} y \rangle } =
        %\frac{dN^i/dy}{<dN/dy>} = 
        \frac{N^i_{\psip}}{N_{\psip}} \, \frac{N^{\rm eq}_{\rm MB}}{N^{i,\rm eq}_{\rm MB}} \,  \frac{(A\epsilon)_{\psip}}{(A\epsilon)^i_{\psip}} \, \frac{\epsilon^i_{\rm MB}}{\epsilon_{\rm MB}} \,,
        \label{eq:relyield}
    \end{equation}
where $N_{\psip}$ is the raw number of $\psip$ signals, $N^{\rm eq}_{\rm MB}$ is the number of equivalent minimum-bias (MB) events (whose calculation is described later in this section), $A\epsilon_{\psip}$ is the average acceptance-times-reconstruction efficiency for $\psip$, and $\epsilon_{\rm MB}$ is the selection efficiency for minimum-bias events. Here the terms denoted with superscript $i$ refer to quantities measured for the multiplicity interval $i$ while terms with no superscript correspond to multiplicity-integrated quantities.

$\Jpsi$ and $\psip$ candidates were reconstructed by forming pairs of opposite-sign-charge muon tracks and computing their invariant mass. Muons were identified by requiring that each track candidate in the MCH matches with a track segment in the MTR. In addition, tracks were required to be reconstructed within $-4.0 < \eta^{\mu} < -2.5$, with their radial distance  from the beam axis at the end of the front absorber, $R_{\rm abs}$, being limited  within $17.6 < R_{\rm abs} < 89.5$ cm, to ensure that only the tracks within detector acceptance were selected for further analysis.
    
\begin{figure}
     \centering
     \subfigure[]{\includegraphics[width=0.49\textwidth]{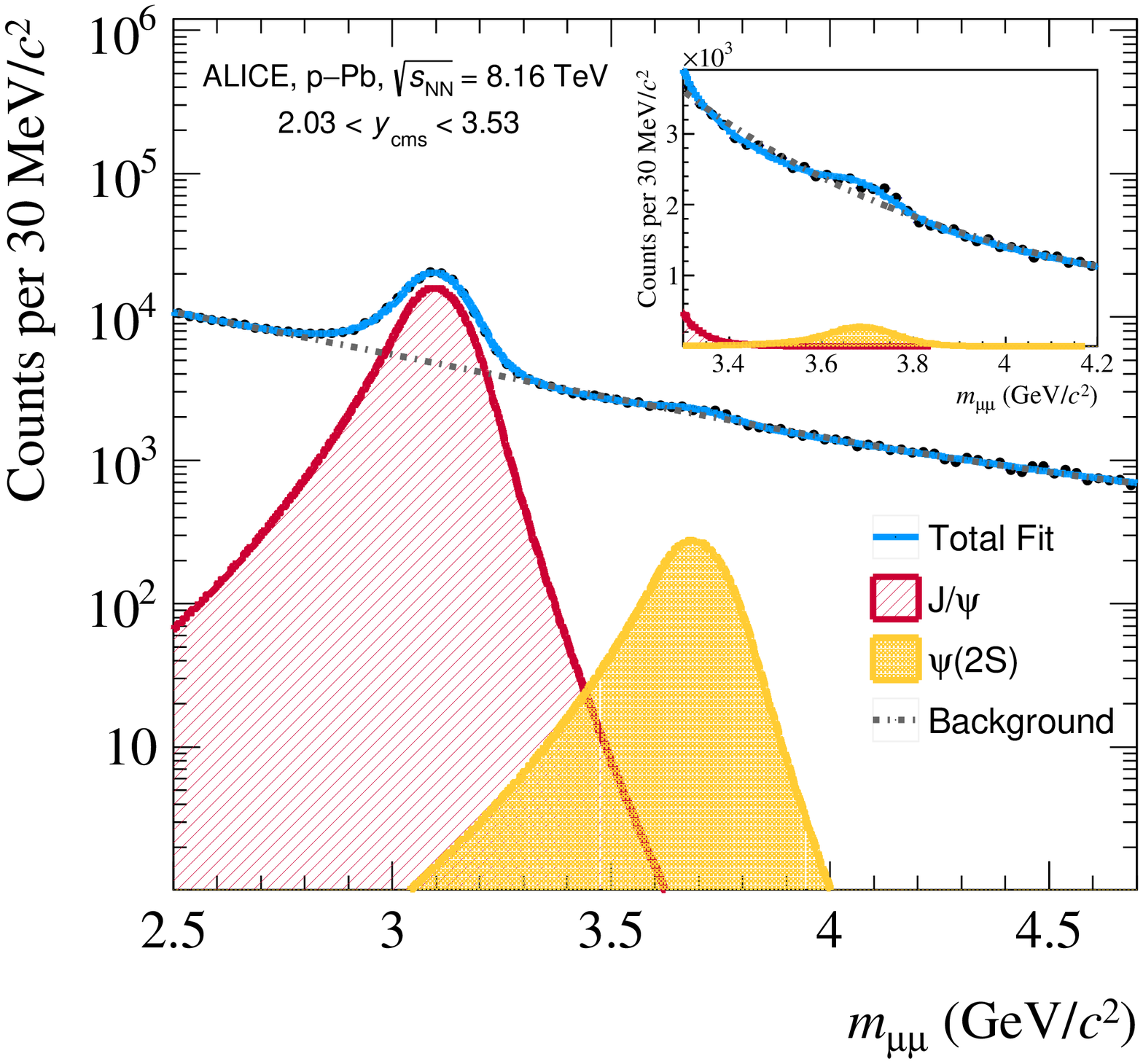}}
     \subfigure[]{\includegraphics[width=0.49\textwidth]{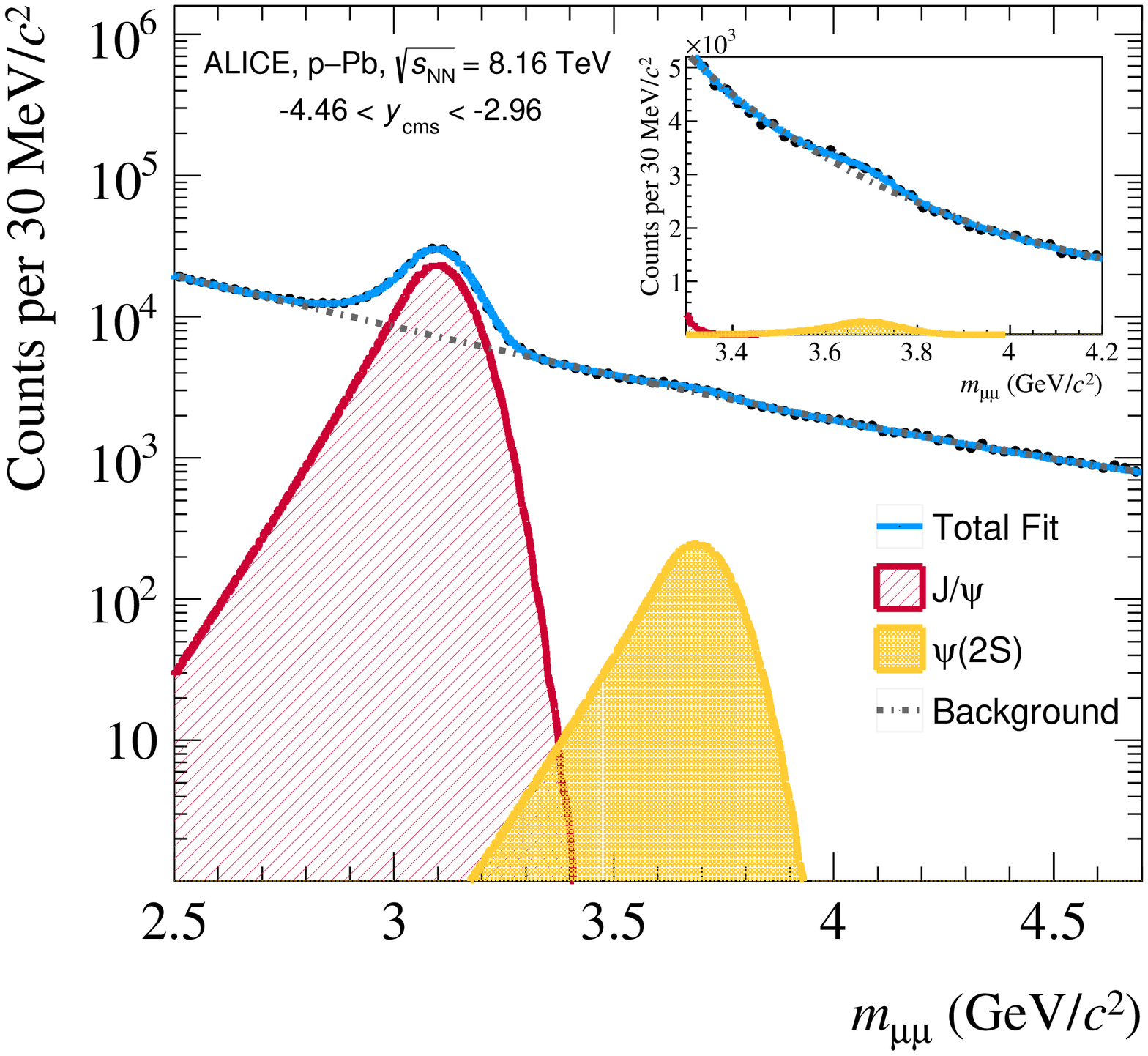}} 
     \subfigure[]{\includegraphics[width=0.49\textwidth]{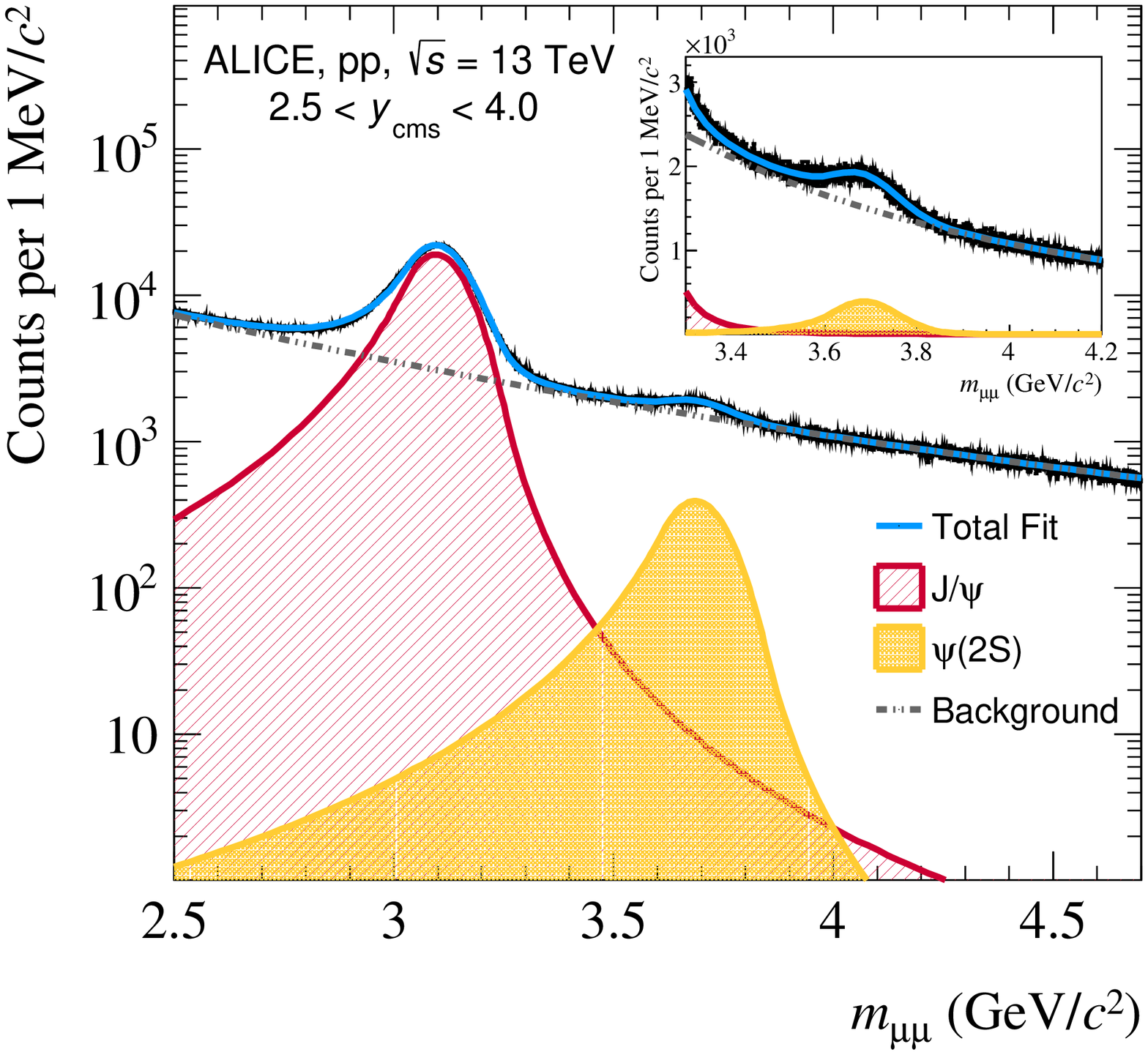}} 
        \caption{Invariant-mass distribution of opposite-sign muon pairs in (a) forward \pPb, (b) backward \pPb, and (c) \pp data, integrated in multiplicity. Superimposed is a typical fit function, shown as a solid blue line (see text on details regarding the fit components). The $\Jpsi$ and $\psip$ contributions are shown as red and yellow shaded areas, respectively. Finally the background is depicted as a dashed gray line. The inset shows the region of the $\psip$ mass.
        \label{fig:invmass}
        }
\end{figure}

  % fitting procedure
Raw charmonium yields were extracted by fitting the invariant-mass distribution with a superposition of $\Jpsi$ and $\psip$ signal functions and a background function. Different combinations of functional forms were used to determine the raw yields and their uncertainties. 
The $\Jpsi$ and $\psip$ signal peaks were parametrized either with a Crystal Ball (CB) function or a pseudo-Gaussian function with power-law tails (as implemented first by the NA60 Collaboration~\cite{ALICE-PUBLIC-2015-006,Shahoyan:2001sd}). 
The two functions differ in their parametrization including a Gaussian core and two asymmetric exponential tails.
The $\Jpsi$ mass, width, and normalization were left free in the fit procedure, while the $\psip$ function parameters, apart from the normalization, were bound to those of $\Jpsi$ as explained in Ref.~\cite{ALICE:2015pgg}. The $\psip$ mass was constrained by requiring that the difference with respect to the $\Jpsi$ mass was the one reported by the Particle Data Group in Ref.~\cite{ParticleDataGroup:2020ssz}. 
The $\psip$ width was taken as proportional to the $\Jpsi$ width. The ratio of $\psip$ and $\Jpsi$ peak widths was determined from Monte Carlo or data, as explained in Ref.~\cite{ALICE:2017leg}, combining the values obtained with these two alternative options to define the raw charmonium yields. 
Tail parameters were obtained from data or Monte Carlo and fixed in the fits. 
In both the \pp and \pPb analyses, the tail parameters extracted from the respective multiplicity-integrated data and MC samples were used in the fit with CB signal function.  For the NA60 function, the tail parametrization could only be extracted from MC for both collision systems. Additionally, the tails extracted from pp data and MC were also considered in the p–Pb analysis as well as those used in analysis described in Ref.~\cite{ALICE:2018mml}. The use of \pp tails in \pPb analysis was motivated by the better determination of the tails in the larger \pp data sample, given the similar experimental conditions for these data-taking periods.
To summarize, in the analysis of \pp collisions, two sets of tail parameters were used in the fit of the CB function and one in the fit with the NA60 function. In the \pPb analysis, a total of five parametrizations of the CB tails and one of the NA60 tails were considered. 
Different functional forms were considered to describe the combinatorial background. For \pp collisions, either a sum of two exponentials or a pseudo-Gaussian function whose width varies linearly with the invariant mass were used. Whereas for \pPb collisions, the functions considered were either the product of an exponential and a fourth-order polynomial, or the sum of two exponential functions. 
The invariant-mass distribution was fitted multiple times per each multiplicity interval, each iteration varying the parametrizations of the signal and background components and varying the fit range. 
The number of candidates and their statistical uncertainties were evaluated from the average of the values over all trials, while their standard deviation was used to determine the systematic uncertainty. 
When computing the normalized charmonium yields or the normalized excited-to-ground state ratio, the systematic uncertainty due to the signal parametrization was considered as correlated across charged-particle multiplicity intervals, while the one due to the background description in the fit was treated as uncorrelated. Example of a fit to the multiplicity-integrated invariant-mass distribution in each dataset is depicted in Fig.~\ref{fig:invmass}.

% normalisation factor 
The equivalent number of MB events, $N^{\rm eq}_{\rm MB}$, was obtained from the number of dimuon-triggered events, $N_{\mumu}$, as $N^{\rm eq}_{\rm MB} = F_{\rm norm} \times N_{\mumu} $, where the normalization factor $F_{\rm norm}$ represents the probability of a dimuon trigger to occur in a MB-triggered event. $F_{\rm norm}$ was evaluated either by computing the probability of a coincidence of these two triggers in data or by exploiting an intermediate single-muon trigger~\cite{ALICE:2018wma}.

%acceptance efficiency 
The acceptance-times-efficiency $A\epsilon$ of charmonia was determined via simulations with the PYTHIA~6 event generator coupled with GEANT3 to transport the particles through the detector. The simulations take into account the variation of the experimental conditions with time. 
As can be derived from Eq.\ref{eq:relyield}, the normalized yield is only sensitive to the variation of $A\epsilon$ with charged-particle multiplicity. No variation of the $A\epsilon$ due to the detector occupancy was observed. % in the analyzed \pp and \pPb data samples.
However, it is sensitive to the possible variation of the $y$ and $\pt$ distributions with multiplicity. The $\Jpsi$ analyses have shown a variation of $\langle \pt \rangle$ with charged-particle multiplicity~\cite{ALICE:2020eji,ALICE:2020msa}. Therefore, the influence of the MC input $y$ and $\pt$ distributions was studied. The exercise was done using an iterative procedure to describe the data for each multiplicity interval. Due to the limited size of the $\psip$ signal, the procedure was performed for $\Jpsi$. 
Three iterations were sufficient to converge, as verified by comparing the simulated $\pt$ and $y$ distributions to those obtained from data and corrected for $A\epsilon$.
For \pp data, a 4\% variation of $A\epsilon$ was found from the lowest to the highest multiplicity interval, and used to correct the normalized yield measurement. No dependence of $A\epsilon$ with charged-particle multiplicity was observed for \pPb collisions, therefore no correction was applied. The different behavior of $A\epsilon$ with charged-particle multiplicity in \pp and \pPb collisions is understood as a consequence of the higher $\langle \pt \rangle$ values in \pp collisions at \mbox{$\sqrt{s} = 13$ TeV}~\cite{ALICE:2017leg}, and the steep increase of the acceptance-times-efficiency with $\pt$ above 3~\GeVc.

    % event selection efficiency : INEL / NSD
The extracted yields needed to be corrected for the efficiency of the MB trigger to select the \mbox{INEL$>0$} (NSD) events, $\epsilon_{{\rm INEL}>0}$ ($\epsilon_{\rm NSD}$). The $\epsilon_{\rm INEL>0}$  ($\epsilon_{\rm NSD})$ was evaluated for each multiplicity interval, as described in Ref.\cite{ALICE:2018wma,ALICE:2020swj}, and found to be equal to unity in all the intervals apart from the one with the least multiplicity.  A 1\% correction was determined and applied to the first multiplicity interval.

The following sources of systematic uncertainty on normalized $\psip$ yields were considered: 
\begin{enumerate*}[label=(\roman*)]
   \item the signal extraction,
   \item the normalization factor,
   \item the event-by-event $\Ntr$ to $\Ntrc$ correction, 
   \item effects of the resolution and of the pile-up on the multiplicity classification, 
   \item the acceptance-times-efficiency correction, and 
   \item the event-selection efficiency.
\end{enumerate*}
Except for the correlated uncertainty on event-selection efficiency and the partly correlated uncertainty on signal extraction, all sources of systematic uncertainty have been considered as uncorrelated in multiplicity.
The systematic uncertainties on the charmonium yields in each multiplicity interval, normalized to their multiplicity-integrated values, were determined directly for the ratios. Likewise, the uncertainties on the normalized excited-to-ground state ratios were determined for the ratio directly rather than propagated from individual state yields.
The details on the signal extraction have been explained previously in this section. The central values were determined averaging the results of the fits by varying the signal and background functions, as well as the invariant-mass fit interval. The systematic uncertainty was evaluated as the standard deviation between the values for all individual trials, ranging from 4\% to 5\% for \pp data, and 4\% to 8\% (5\% to 13\%) for \pPb data at forward (backward) rapidity. 
The influence of the normalization factor was evaluated by computing $F_{\rm norm}$ with different methods and studying its variation with time (see Ref.~\cite{ALICE:2020eji}). The effect on the result was found to be negligible in the measured multiplicity intervals. 
The impact of the chosen method for the event-by-event $\Ntr$ to $\Ntrc$ correction was studied in Refs.~\cite{ALICE:2020eji,ALICE:2012pet,ALICE:2015ikl}. Both the reference value used to scale the $\Ntr$ distribution as well as the randomization function considered to introduce a Poissonian fluctuation of the values were varied. The influence of these variations on the normalized yields ranges around 1--2\% (2\%) for \pp (\pPb) collisions. 
The influence of the resolution on the multiplicity axis and the possible remaining pile-up were evaluated as a single contribution by repeating the analysis multiple times by varying the pile-up rejection criteria or introducing a small shift of the $\Ntrc$ intervals. The estimated uncertainty amounts to 2\% in \pPb results and was found to be negligible for \pp data.  
The determination of the acceptance-times-efficiency correction has been described previously in this section. 
The $A\epsilon$ of $\Jpsi$ and $\psip$ were compared as a function of $\pt$, and found to have a similar behavior around their $\langle\pt\rangle$ values. 
To account for possible remaining differences between the variation of the $\Jpsi$ and $\psip$ $A\epsilon$, a conservative uncertainty was assigned to each point, which is as large as the correction (up to 4\% for \pp, negligible for \pPb). 
The uncertainty on event-class normalization, originating from the INEL$>0$ or NSD event-selection efficiency, was evaluated as in Ref.~\cite{ALICE:2018wma}. A 1\% uncertainty was assigned to both \pp and \pPb measurements, correlated in multiplicity. 
The contributions to the systematic uncertainty of the normalized $\psip$ yields are summarized in Table~\ref{tab:sys_psipY}.

    \begin{table}[t]
        \centering
        \caption{
        \label{tab:sys_psipY}
        Sources of systematic uncertainty on the normalized $\psip$ yields. The values marked with asterisk (diamond) are (partly) correlated in multiplicity.
        }
        \begin{tabular}{l|c|c c}
        \hline 
         Source & \pp & \multicolumn{2}{c}{\pPb}\\
        
         & $2.5 < \ycms < 4.0$ & $2.03 < \ycms < 3.53$ & $-4.46 < \ycms < -2.96$ \\
        \hline 
        Signal extraction           & 4--5\% $\diamond$  & 4--8\% $\diamond$ & 5--13\% $\diamond$ \\
        Event-by-event $\Ntrc$      & 1--2\%    & 2\%   & 2\% \\
        Resolution and pile-up &  --         & 2\%      & 2\% \\
        Acceptance-times-efficiency & 0-4\% $\diamond$ & -- & -- \\
        Event-class normalization   & 1\%*       & 1\%*     & 1\%* \\
        \hline 
        \end{tabular}
    \end{table}

The same sources of systematic uncertainty were studied for the normalized $\psipOJpsi$ ratios. In this case, the influence of the event-by-event $\Ntrc$ correction, as well as that of the resolution and pile-up were found to be negligible. The uncertainty originating from the normalization to the INEL$>0$ (NSD) event class in \pp (\pPb) collisions cancels out in the ratio. 
Signal extraction is the sole contributor to the systematic uncertainty on the normalized $\psipOJpsi$ ratio, and amounts to 4--5\% for \pp results, and to 4--8\% (5--12\%) for forward (backward) rapidity \pPb measurements.

\section{Results}
\label{sec:Results}

In \pp collisions, the measured charged-particle multiplicity spans up to about six times the average value. In \pPb collisions at forward (backward) rapidity, the yields have been measured up to about three (four) times the average multiplicity.
The normalized $\psip$ yield increases with increasing multiplicity, presenting a similar trend in \pp collisions at \thirteen and \pPb collisions at \eightnn. 
The \pPb results at forward and backward rapidity are compatible between each other within uncertainties. 

\begin{figure}
    \centering
    \includegraphics[width=0.65\textwidth]{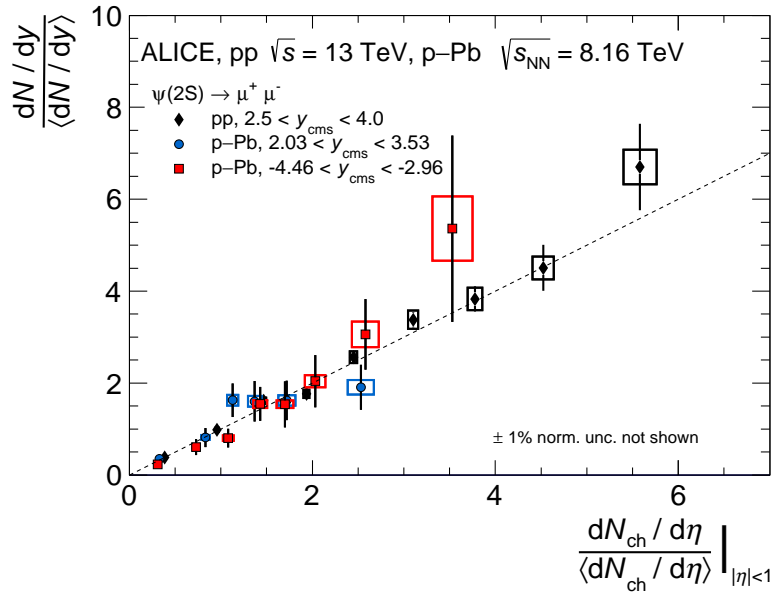}
    \caption{
    The normalized $\psip$  yield as a function of the normalized charged-particle pseudorapidity density in \pp collisions at \thirteen and \pPb collisions at \eightnn. Quoted is the correlated event-class normalization uncertainty. 
    The \pp results are normalized to the INEL$>$0 event class, whereas the \pPb ones refer to the NSD one. 
    \label{fig:yield}
    }
\end{figure}

\begin{figure}
    \centering
    \includegraphics[width=0.65\textwidth]{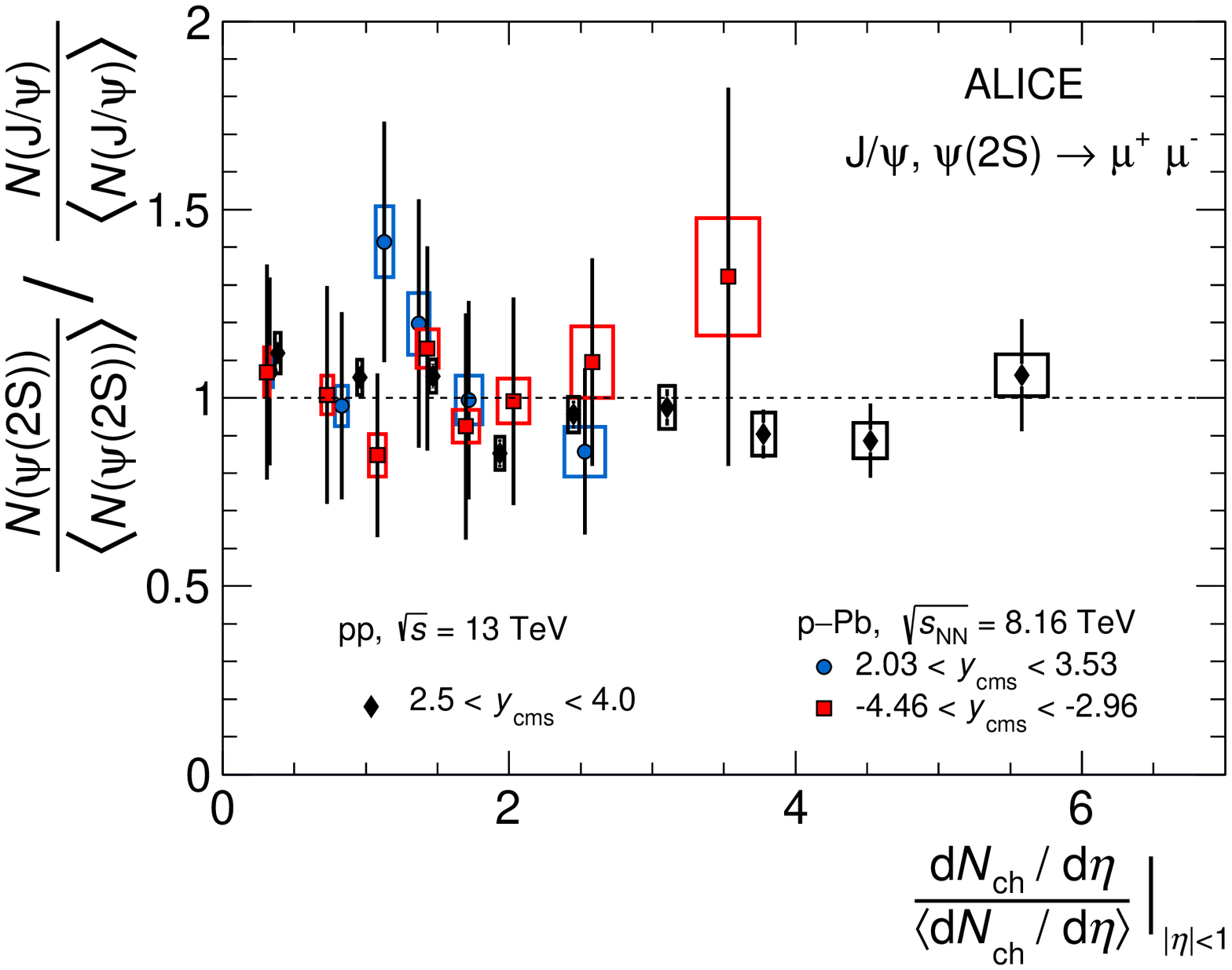}
    \caption{
    Ratio of normalized $\psipOJpsi$ yield as a function of the normalized charged-particle pseudorapidity density in \pp collisions at \thirteen and \pPb collisions at \eightnn.
    The \pp results are normalized to the INEL$>$0 event class, whereas the \pPb ones refer to the NSD one. 
    \label{fig:2sO1s}
    }
\end{figure}

\begin{figure}[!htbp]
 \centering
    \subfigure[Normalized $\psip$ yield]{
    \includegraphics[width =0.45\textwidth]{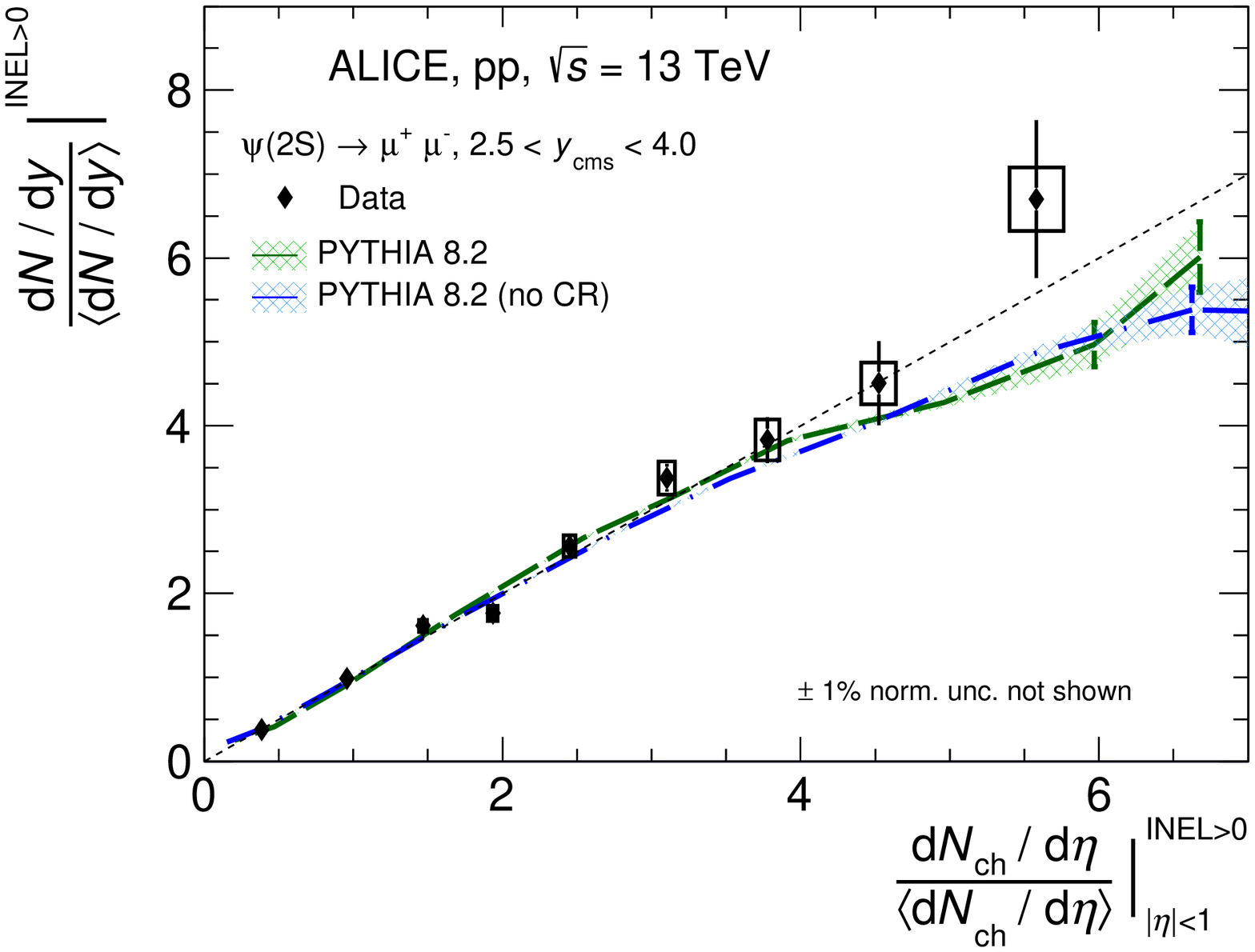}
    \label{fig:ModelPPYield}
  }
    \qquad 
    \subfigure[Normalized ratio of $\psip$ over $\Jpsi$]{
    \includegraphics[width =0.45\textwidth]{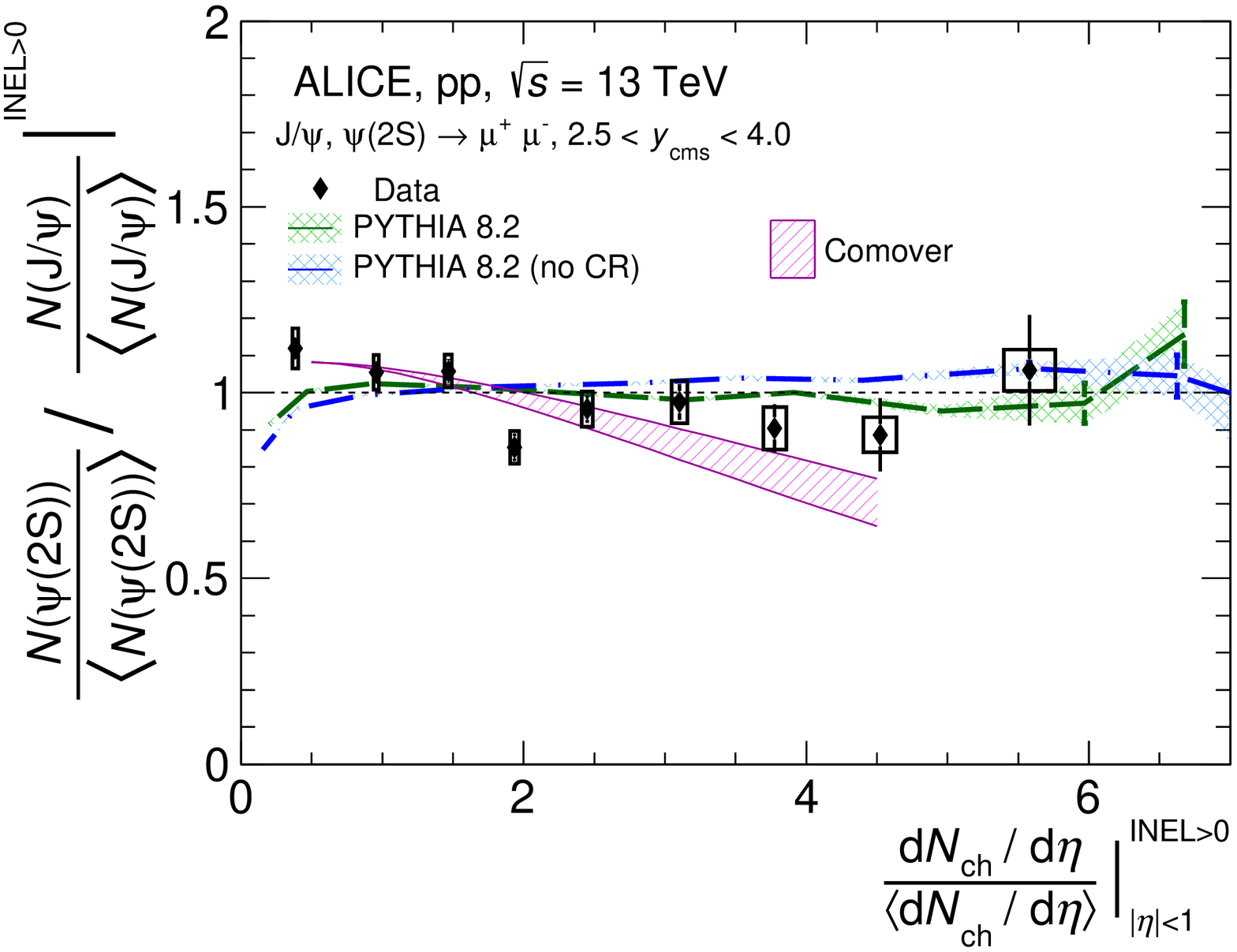}
    \label{fig:ModelPPRatio}
  }
    \caption{
    The normalized $\psip$ yield and the ratio of normalized $\psipOJpsi$ yields at $2.5 < \ycms <4.0$ as a function of the normalized charged-particle pseudorapidity density in \pp collisions at \thirteen. Measurements are compared with the following models: PYTHIA~8.2~\cite{Sjostrand:2014zea}, PYTHIA~8.2 without color reconnection (no CR)~\cite{Sjostrand:2014zea}, comovers~\cite{Ferreiro:2014bia}. Quoted is the correlated event-class normalization uncertainty.
    \label{fig:ModelPP}
    }
\end{figure}

\begin{figure}[!htbp]
 \centering
     \subfigure[Normalized $\psip$ yield]{
     \includegraphics[width =0.45\textwidth]{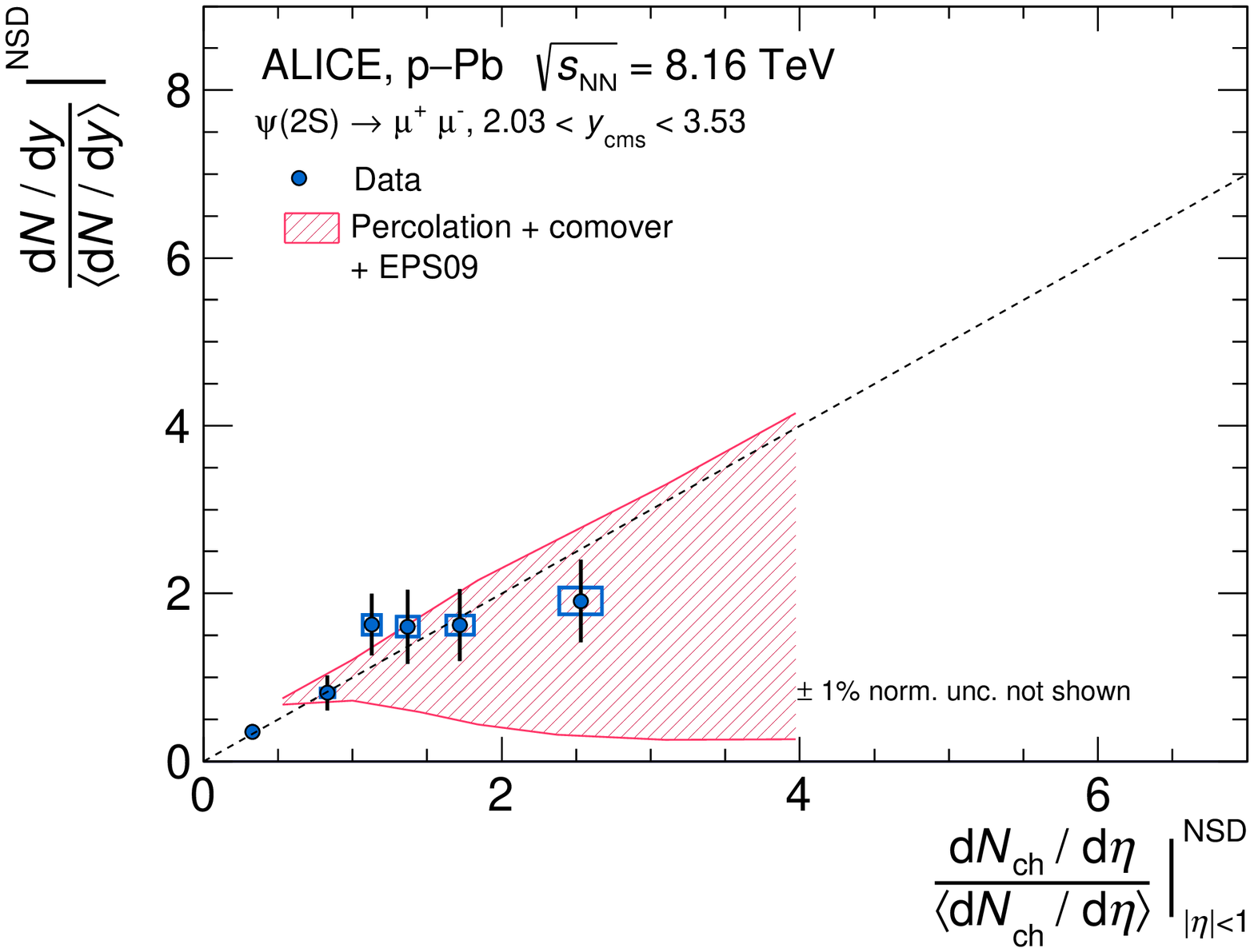}
     \label{fig:ModelPPbYieldFw}
   }
     \qquad 
     \subfigure[Normalized ratio of $\psip$ over $\Jpsi$]{
    \includegraphics[width =0.45\textwidth]{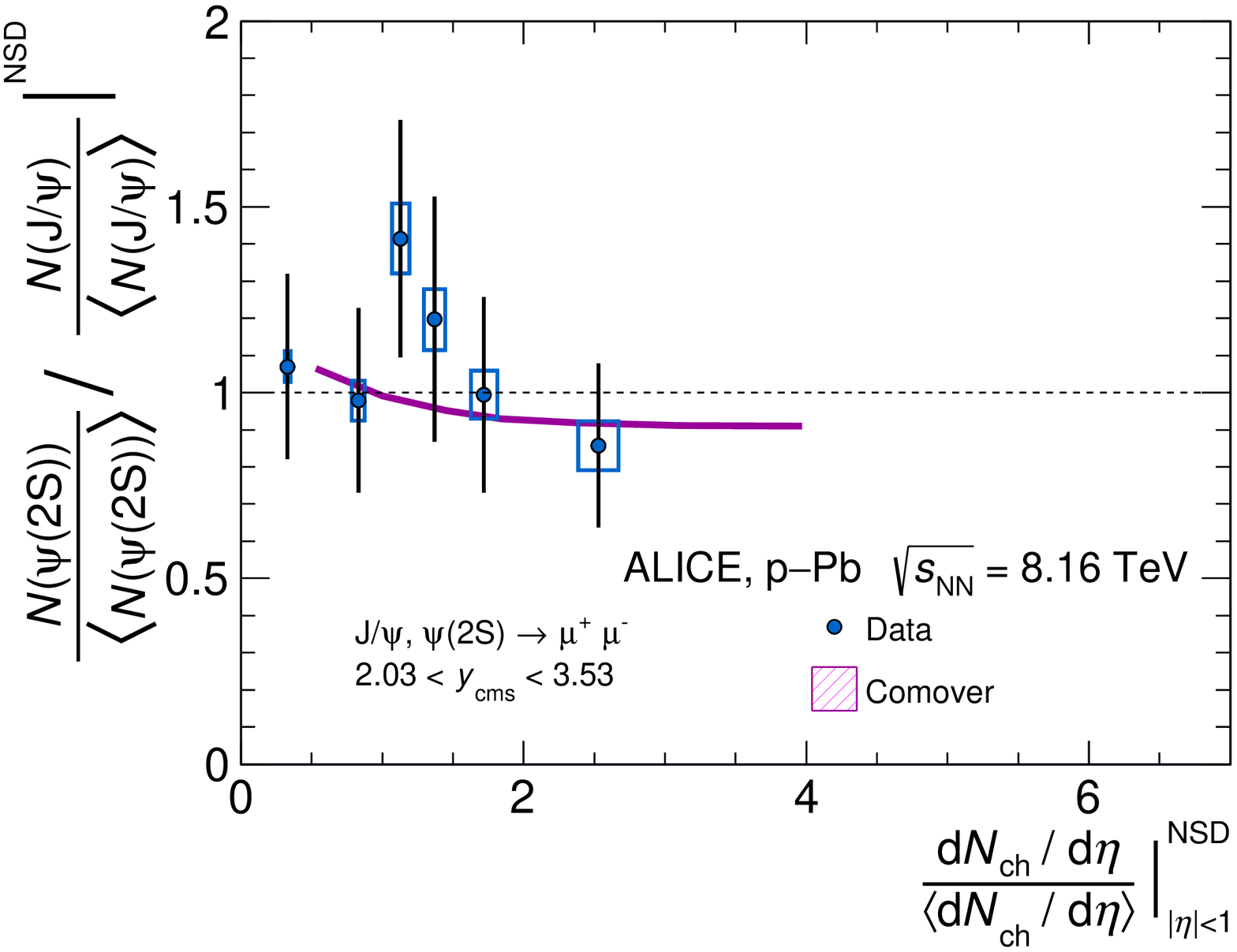}
     \label{fig:ModelPPbRatioFw}
   }
   
    \caption{
    The normalized $\psip$ yield and the ratio of normalized $\psipOJpsi$ yields at $2.03<\ycms<3.53$ as a function of the normalized charged-particle pseudorapidity density in \pPb collisions at \eightnn. Measurements are compared with 
    the percolation calculation~\cite{Armesto:1996kt} coupled with comover model~\cite{Ferreiro:2014bia} and EPS09 nPDF~\cite{Eskola:2009uj}. Quoted is the correlated event-class normalization uncertainty.
    \label{fig:ModelPPbFw}
    }
\end{figure}

\begin{figure}[!htbp]
 \centering
     \subfigure[Normalized $\psip$ yield]{
    \includegraphics[width =0.45\textwidth]{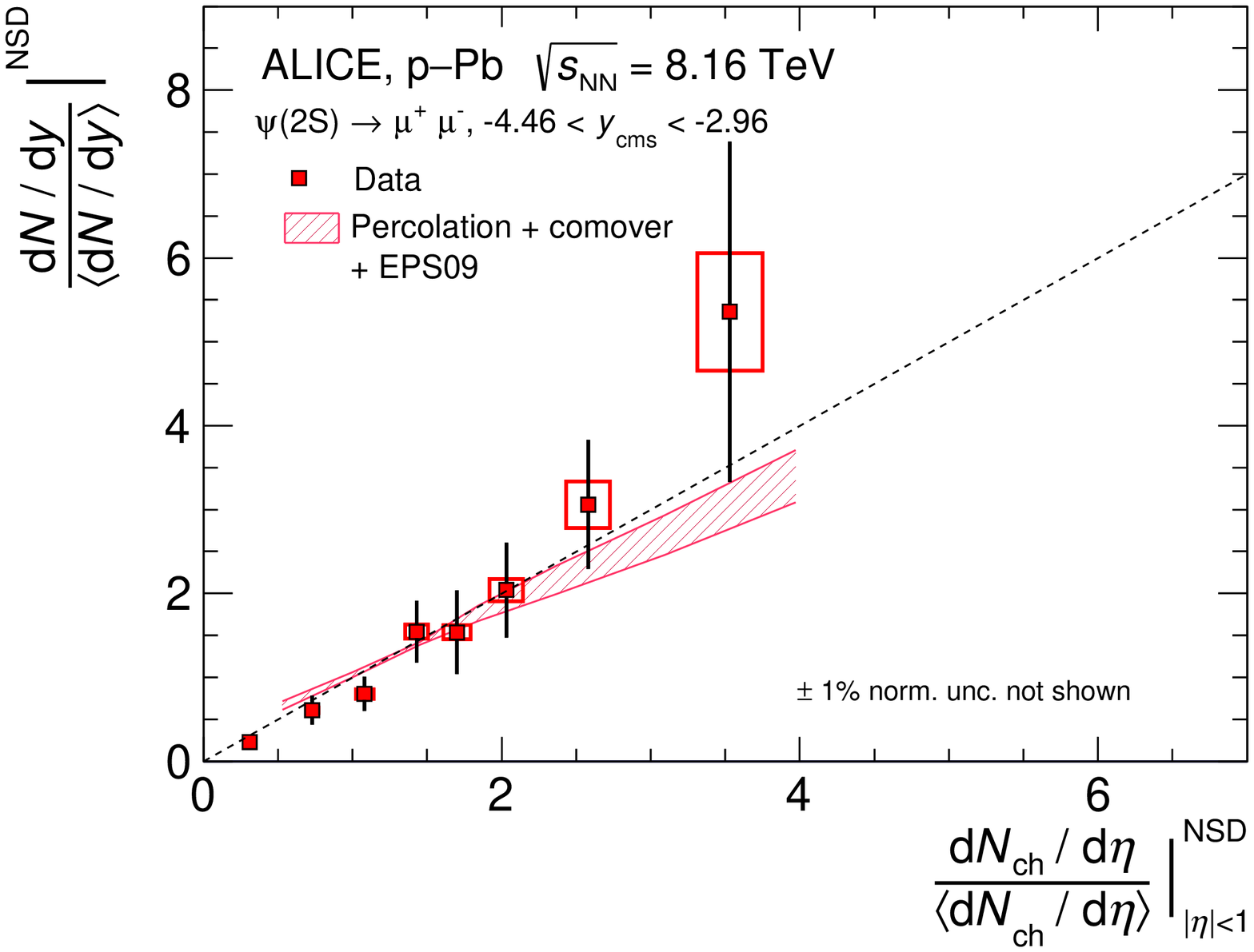}
     \label{fig:ModelPPbYieldBw}
   }
     \qquad 
     \subfigure[Normalized ratio of $\psip$ over $\Jpsi$]{
    \includegraphics[width =0.45\textwidth]{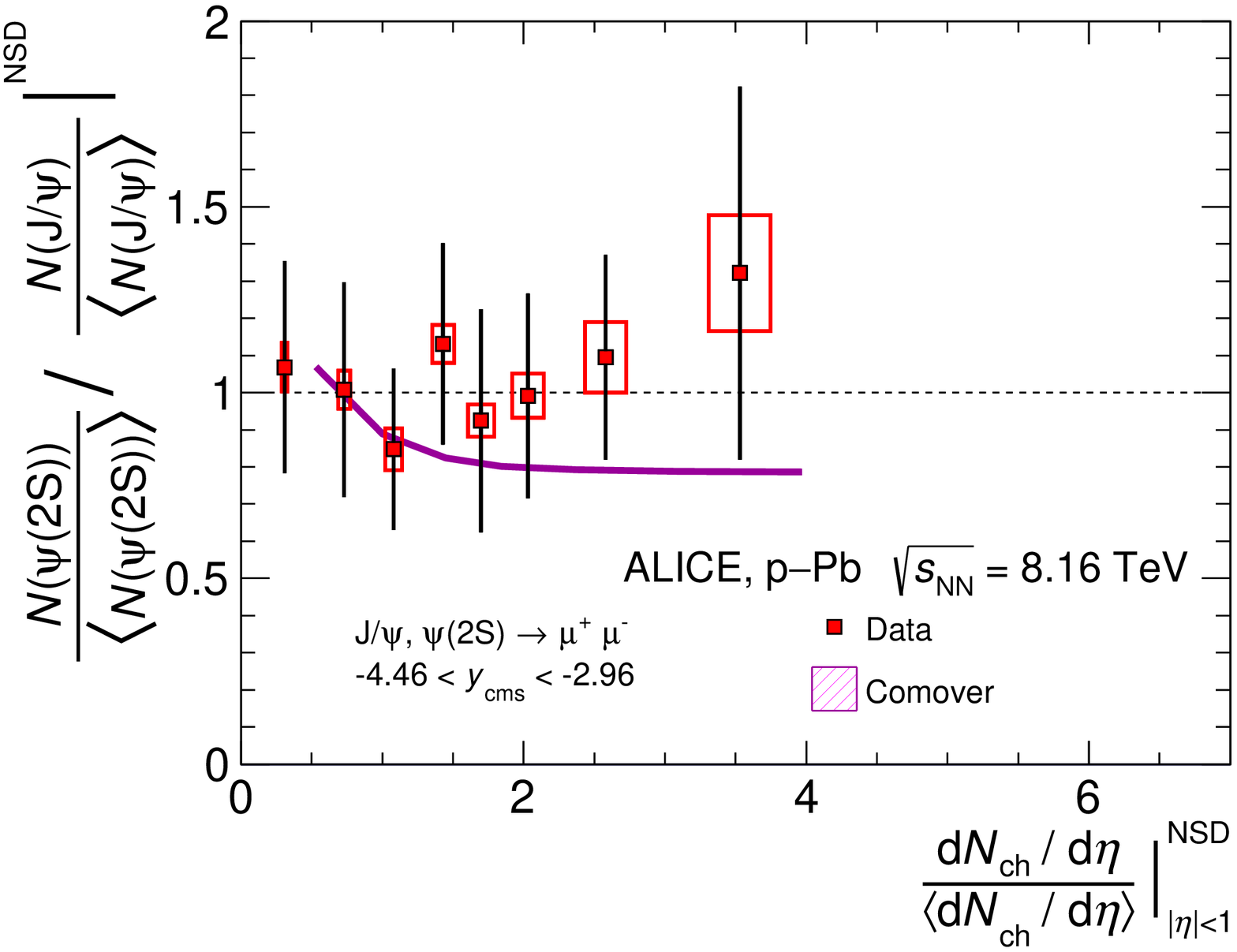}
     \label{fig:ModelPPbRatioBw}
   }
    \caption{
    The normalized $\psip$ yield and the ratio of normalized $\psipOJpsi$ yields at $-4.46<\ycms<-2.96$ as a function of the normalized charged-particle pseudorapidity density in \pPb collisions at \eightnn. Measurements are compared with 
    the percolation calculation~\cite{Armesto:1996kt} coupled with comover model~\cite{Ferreiro:2014bia} and EPS09 nPDF~\cite{Eskola:2009uj}. Quoted is the correlated event-class normalization uncertainty.
    \label{fig:ModelPPbBw}
    }
\end{figure}

The ratio of normalized $\psipOJpsi$ yields is evaluated to outline possible differences between multiplicity dependence of the production of the excited $\psip$ and ground-state $\Jpsi$ with reduced uncertainties. 
The double ratios, obtained by dividing the $\psipOJpsi$ yield ratios in multiplicity intervals by the multiplicity-integrated ratios, are shown in Fig.~\ref{fig:2sO1s}.
The measurements are compatible with unity within uncertainties for both colliding systems. 
The double ratio in \pp collisions, which is more precise than that in \pPb, is consistent with a linear trend, either with a null ($\chi^2/$ndf = 2.1) or a negative slope ($\chi^2/$ndf = 1.5, free value of slope $-0.05 \pm 0.02$).
The measurements are compared with theoretical calculations in Fig.~\ref{fig:ModelPP},~\ref{fig:ModelPPbFw}, and~\ref{fig:ModelPPbBw}. 

%
% PYTHIA 8
%
The results from the PYTHIA~8.2 event generator~\cite{Sjostrand:2014zea} for \pp collisions at $\sqrt{s} = 13$ TeV are shown in Fig.\ref{fig:ModelPP}. 
In contrast to PYTHIA~6, PYTHIA~8.2 allows charm and beauty quarks to be involved in secondary hard processes, i.e.,\ in multiparton interactions (MPI). 
From the implementation of the MPI mechanism a simple scaling is expected, in which the multiplicity of charged particles is proportional to the number of MPI and to the amount of hard processes taking place in a collision. At first order, this results in an increasing trend of the normalized quarkonium yields as a function of the normalized charged-particle multiplicity, with a slope close to unity. 
The $\psip$ yield measurement is described within uncertainties by the PYTHIA~8.2 event generator, both with and without the color-reconnection (CR) contribution. 
No significant difference is observed between the two configurations of PYTHIA 8.2, even though the naive expectation is that of a steeper trend in the simulation with CR, caused by the effective reduction of the charged-particle multiplicity at large multiplicities in the CR scenario. 
A tension appears on the comparison of the measurement and the calculation of the $\psipOJpsi$ ratio as a function of the charged-particle multiplicity, as can be seen in Fig.\ref{fig:ModelPPRatio}. For values of normalized charged-particle multiplicity below unity, the simulations yield values of the $\psipOJpsi$ ratio lower than unity, while the measured values in the same multiplicity range reach above unity. 
A different event activity bias could be the explanation for the discrepancy found at low multiplicity. 
For instance, in PYTHIA 8.2, events with $\psip$ are, on average, biased towards a larger event activity as a consequence of its larger mass. This interpretation can not be confirmed, nor refuted, with the current measurement precision.

%
% EPOS 3
% 

% 
% Percolation
% 
High-energy hadronic collisions can also be simulated in the percolation framework~\cite{Armesto:1996kt}. This paper shows calculations from this model only for \pPb collisions. The key ingredient of this model are the color ropes or flux tubes (strings), that are formed in each parton--parton interaction and constitute the main source of particle production. The strings have non-negligible transverse size and can interact among each other. In particular, they can overlap, reducing their effective number and, consequently, the particle production.
In this string model~\cite{Ferreiro:2012fb}, the number of quarkonia is assumed to be proportional to the number of partonic scatterings, which corresponds to the number of produced strings. Instead, the charged-particle multiplicity scales with the number of participants due to the influence of shadowing~\cite{Ferreiro:2008wc}, parton saturation~\cite{Kharzeev:2008nw}, or percolation~\cite{Armesto:1996kt}. 
For \pPb collisions, the percolation calculations shown in this paper are coupled to the comover model~\cite{Ferreiro:2014bia,Esposito:2020ywk} and to EPS09 nPDFs~\cite{Eskola:2009uj}, the latter having been added to account for nuclear effects.
In the comover model~\cite{Ferreiro:2014bia,Esposito:2020ywk}, quarkonia can be dissociated by interacting with the surrounding comoving particles in the final state. The probability for this to happen depends on the binding energy of each quarkonium state and on the density of comoving particles. 
The latter determines the uncertainties of the model. The rapidity distribution of the density of hadrons is parametrized taking into account the geometry of the collision~\cite{Armesto:1998rc}. 
The EPS09 uncertainties have a sizable influence on the model estimate of the yields but cancel in the $\psipOJpsi$ ratio. 
Feed-down contributions from decays of other charmonium states are taken into account in the calculation. 
The normalized $\psip$ yields in \pPb collisions are described within   uncertainties by the percolation + comover + EPS09 calculation (see Fig.\ref{fig:ModelPPbYieldFw} and~\ref{fig:ModelPPbYieldBw}). 
The model expects a nearly linear increase of the yield at backward rapidity, compatible with the measured one. The model calculation at forward rapidity also describes the measurement, within the large EPS09 uncertainties. It is to be noted that the measured values are on the upper edge of the uncertainty band of the prediction.

In the comover scenario~\cite{Ferreiro:2014bia,Esposito:2020ywk}, the probability of $\psip$ to dissociate due to interactions in the final state is larger than that of $\Jpsi$ due to its lower binding energy. The effect increases with charged-particle multiplicity, i.e., with the comover density. This results in a decreasing trend of the double ratio with increasing charged-particle multiplicity, in contrast to PYTHIA~8.2 simulations for \pp collisions, which do not include final-state effects. 
The uncertainties shown in the comover calculation for the double ratio in \pp collisions represent the influence of varying by $15\%$ the density of comoving particles. 
The double ratios for \pp collisions are described by the comover calculation within uncertainties, see Fig.\ref{fig:ModelPPRatio}. 
The data-to-model comparison suggests a steeper decrease in the model calculation than in the measured data points, albeit no firm conclusion can be drawn with the current precision of the measurement.
The double ratios measured in \pPb collisions are shown in Figs.~\ref{fig:ModelPPbRatioFw} and~\ref{fig:ModelPPbRatioBw}. The measurements, while weighted by a large uncertainty, are consistent with the comover calculation, which predicts a stronger suppression of the excited states at backward rapidity. Previous studies of the relative suppression between the two charmonium states indeed revealed a stronger suppression at backward rapidity, largely independent of multiplicity~\cite{ALICE:2015kgk,ALICE:2020tsj}. 
%%%%

To better contextualize the results presented in this paper, one needs to consider previously published results on multiplicity-dependent quarkonium production.
The normalized yields of charmonium and bottomonium ground and excited states at large rapidity increase with the normalized charged-particle multiplicity at midrapidity with a similar approximately linear trend (with gradient equal to unity) in \pp collisions~\cite{ALICE:2020eji,ALICE:2021zkd,ALICE:UpsvsMult:2022}. 
A steeper increase is observed for $\Jpsi$ production at midrapidity~\cite{ALICE:2015ikl,ALICE:2020msa}. 
All models for $\Jpsi$ production in \pp collisions at midrapidity (PYTHIA 8.2, model with coherent particle production (CPP)~\cite{Kopeliovich:2013yfa}, EPOS3~\cite{Werner:2013tya}, Color Glass Condensate effective theory (CGC)~\cite{Ma:2018bax}, 3-pomeron CGC~\cite{Levin:2019fvb}, and percolation~\cite{Ferreiro:2012fb}) predict a faster-than-linear increase of the yields with charged-particle pseudorapidity density at midrapidity~\cite{ALICE:2020msa}. All these models predict an effective reduction of the charged-particle multiplicity at high multiplicities due to different physics mechanisms (color reconnection, coherent particle production,  3-gluon fusion, saturation, or percolation). The percolation model slightly overestimates the yield at high multiplicity, while PYTHIA~8.2 and EPOS3 underpredict the data. The CPP, CGC, and 3-pomeron CGC models give a good description of the measurements. 
To interpret these results, one should keep in mind that in all models, except PYTHIA, only the prompt component is considered, and the non-prompt contribution exhibits a stronger increase than the prompt one with multiplicity in PYTHIA~8.2~\cite{ALICE:2020msa}. 
Model calculations  predict a slightly smaller increase for $\Jpsi$ yields at large rapidity than at midrapidity with respect to charged-particle pseudorapidity density at midrapidity, consistent with the measurements. This suggests that the influence of the physics mechanisms at play differs for large and midrapidity $\Jpsi$.  
The measurements presented in this paper for $\psip$ and in Ref.~\cite{ALICE:UpsvsMult:2022} for bottomonium ground and excited states at large rapidity are in agreement with this picture. 
It should be noted that, despite the recent progress, there are not many predictions available for excited charmonium states or bottomonium states as a function of charged-particle multiplicity.

\section{Summary}
\label{sec:Summary}

The first measurements of $\psip$ production and of the $\psipOJpsi$ production ratio as a function of charged-particle multiplicity in \pp and \pPb collisions at the LHC are presented. 
Charmonium yields were measured at large rapidity, whereas charged-particle multiplicity was measured at central rapidity. 
Both charmonium yields and the charged-particle multiplicity have been normalized to their respective multiplicity-integrated values. 
The normalized $\psip$ yield increases with the normalized charged-particle density in both collision systems with an approximately linear trend with slope close to unity. 
The normalized $\psipOJpsi$ yield ratio is compatible with unity independently of the charged-particle multiplicity within uncertainties, suggesting a similar multiplicity dependence for excited and ground state charmonium states. 
The results can be described by models (PYTHIA~8.2 , percolation + comover + EPS09) within uncertainties. 

The measurements of the charged-particle multiplicity dependence of excited-to-ground state ratios provide additional constraints to models. 
The results for charmonium and bottomonium states at large rapidity in \pp and \pPb collisions provide a coherent picture, with ratios compatible with unity. The precision of the measurements does not allow one to rule out neither the decrease with increasing charged-particle multiplicity predicted by the comover model, nor the nearly flat trend of PYTHIA~8.2 calculations. 
Bottomonium excited-to-ground state ratios at midrapidity show a smooth decreasing trend with increasing charged-particle pseudorapidity density at midrapidity from \pp to \pPb and to \PbPb collisions~\cite{CMS:2013jsu,CMS:2020fae}. A data-to-model comparison is unfortunately missing for the latter.

%%%%%%%%%%%%%%%%%%%%%%%%%%%%%%%%
% end main text 
%%%%%%%%%%%%%%%%%%%%%%%%%%%%%%%%

%%%%% acknowledgements - handled by EB chairs 
\newenvironment{acknowledgement}{\relax}{\relax}
\begin{acknowledgement}
\section*{Acknowledgements}
% add specific acknowledgements here 
% ...but please don't remove the line below: funding agencies
% will be acknowledged with a custom tex file handled by EB chairs after Collab Round 2
We gratefully acknowledge E. G. Ferreiro and B. Kopeliovich for instructive discussions, and E. G. Ferreiro for providing the theoretical calculations. 
% Version: 2022-03-14

The ALICE Collaboration would like to thank all its engineers and technicians for their invaluable contributions to the construction of the experiment and the CERN accelerator teams for the outstanding performance of the LHC complex.
The ALICE Collaboration gratefully acknowledges the resources and support provided by all Grid centres and the Worldwide LHC Computing Grid (WLCG) collaboration.
The ALICE Collaboration acknowledges the following funding agencies for their support in building and running the ALICE detector:
A. I. Alikhanyan National Science Laboratory (Yerevan Physics Institute) Foundation (ANSL), State Committee of Science and World Federation of Scientists (WFS), Armenia;
Austrian Academy of Sciences, Austrian Science Fund (FWF): [M 2467-N36] and Nationalstiftung f\"{u}r Forschung, Technologie und Entwicklung, Austria;
Ministry of Communications and High Technologies, National Nuclear Research Center, Azerbaijan;
Conselho Nacional de Desenvolvimento Cient\'{\i}fico e Tecnol\'{o}gico (CNPq), Financiadora de Estudos e Projetos (Finep), Funda\c{c}\~{a}o de Amparo \`{a} Pesquisa do Estado de S\~{a}o Paulo (FAPESP) and Universidade Federal do Rio Grande do Sul (UFRGS), Brazil;
Bulgarian Ministry of Education and Science, within the National Roadmap for Research Infrastructures 2020-2027 (object CERN), Bulgaria;
Ministry of Education of China (MOEC) , Ministry of Science \& Technology of China (MSTC) and National Natural Science Foundation of China (NSFC), China;
Ministry of Science and Education and Croatian Science Foundation, Croatia;
Centro de Aplicaciones Tecnol\'{o}gicas y Desarrollo Nuclear (CEADEN), Cubaenerg\'{\i}a, Cuba;
Ministry of Education, Youth and Sports of the Czech Republic, Czech Republic;
The Danish Council for Independent Research | Natural Sciences, the VILLUM FONDEN and Danish National Research Foundation (DNRF), Denmark;
Helsinki Institute of Physics (HIP), Finland;
Commissariat \`{a} l'Energie Atomique (CEA) and Institut National de Physique Nucl\'{e}aire et de Physique des Particules (IN2P3) and Centre National de la Recherche Scientifique (CNRS), France;
Bundesministerium f\"{u}r Bildung und Forschung (BMBF) and GSI Helmholtzzentrum f\"{u}r Schwerionenforschung GmbH, Germany;
General Secretariat for Research and Technology, Ministry of Education, Research and Religions, Greece;
National Research, Development and Innovation Office, Hungary;
Department of Atomic Energy Government of India (DAE), Department of Science and Technology, Government of India (DST), University Grants Commission, Government of India (UGC) and Council of Scientific and Industrial Research (CSIR), India;
National Research and Innovation Agency - BRIN, Indonesia;
Istituto Nazionale di Fisica Nucleare (INFN), Italy;
Japanese Ministry of Education, Culture, Sports, Science and Technology (MEXT) and Japan Society for the Promotion of Science (JSPS) KAKENHI, Japan;
Consejo Nacional de Ciencia (CONACYT) y Tecnolog\'{i}a, through Fondo de Cooperaci\'{o}n Internacional en Ciencia y Tecnolog\'{i}a (FONCICYT) and Direcci\'{o}n General de Asuntos del Personal Academico (DGAPA), Mexico;
Nederlandse Organisatie voor Wetenschappelijk Onderzoek (NWO), Netherlands;
The Research Council of Norway, Norway;
Commission on Science and Technology for Sustainable Development in the South (COMSATS), Pakistan;
Pontificia Universidad Cat\'{o}lica del Per\'{u}, Peru;
Ministry of Education and Science, National Science Centre and WUT ID-UB, Poland;
Korea Institute of Science and Technology Information and National Research Foundation of Korea (NRF), Republic of Korea;
Ministry of Education and Scientific Research, Institute of Atomic Physics, Ministry of Research and Innovation and Institute of Atomic Physics and University Politehnica of Bucharest, Romania;
Ministry of Education, Science, Research and Sport of the Slovak Republic, Slovakia;
National Research Foundation of South Africa, South Africa;
Swedish Research Council (VR) and Knut \& Alice Wallenberg Foundation (KAW), Sweden;
European Organization for Nuclear Research, Switzerland;
Suranaree University of Technology (SUT), National Science and Technology Development Agency (NSTDA), Thailand Science Research and Innovation (TSRI) and National Science, Research and Innovation Fund (NSRF), Thailand;
Turkish Energy, Nuclear and Mineral Research Agency (TENMAK), Turkey;
National Academy of  Sciences of Ukraine, Ukraine;
Science and Technology Facilities Council (STFC), United Kingdom;
National Science Foundation of the United States of America (NSF) and United States Department of Energy, Office of Nuclear Physics (DOE NP), United States of America.
In addition, individual groups or members have received support from:
Marie Sk\l{}odowska Curie, Strong 2020 - Horizon 2020, European Research Council (grant nos. 824093, 896850, 950692), European Union;
Academy of Finland (Center of Excellence in Quark Matter) (grant nos. 346327, 346328), Finland;
Programa de Apoyos para la Superaci\'{o}n del Personal Acad\'{e}mico, UNAM, Mexico.

\end{acknowledgement}

%%%%%%%% Bibliography 
\bibliographystyle{utphys}   % Remember we use title in the biblio
%\bibliography{bibliography}
\bibliography{bibliography,alice_papers,bibliography_intro,bibliography_detector,bibliography_results,bibliography_multiplicity}
 %\input {bibliography.tex}  

%%%%%%%%%%%%%%%%%%%%%%%%%%%%%%%%
% Appendices: yours (if any) + authorlist
%%%%%%%%%%%%%%%%%%%%%%%%%%%%%%%%
\newpage
\appendix

%
%\input{} % put your appendices here (if any)
%

%%%%% Authorlist - please do not touch: handled by EB chairs 
\section{The ALICE Collaboration}
\label{app:collab}
% ALICE Collaboration author list for 2022-03-14
\begin{flushleft} 
\small

S.~Acharya\,\orcidlink{0000-0002-9213-5329}\,$^{\rm 124,131}$, 
D.~Adamov\'{a}\,\orcidlink{0000-0002-0504-7428}\,$^{\rm 86}$, 
A.~Adler$^{\rm 69}$, 
G.~Aglieri Rinella\,\orcidlink{0000-0002-9611-3696}\,$^{\rm 32}$, 
M.~Agnello\,\orcidlink{0000-0002-0760-5075}\,$^{\rm 29}$, 
N.~Agrawal\,\orcidlink{0000-0003-0348-9836}\,$^{\rm 50}$, 
Z.~Ahammed\,\orcidlink{0000-0001-5241-7412}\,$^{\rm 131}$, 
S.~Ahmad\,\orcidlink{0000-0003-0497-5705}\,$^{\rm 15}$, 
S.U.~Ahn\,\orcidlink{0000-0001-8847-489X}\,$^{\rm 70}$, 
I.~Ahuja\,\orcidlink{0000-0002-4417-1392}\,$^{\rm 37}$
% ...

\section*{Affiliation Notes}

$^{\rm I}$ Deceased\\
$^{\rm II}$ Also at: Max-Planck-Institut f\"{u}r Physik, Munich, Germany\\
$^{\rm III}$ Also at: Italian National Agency for New Technologies, Energy and Sustainable Economic Development (ENEA), Bologna, Italy\\
$^{\rm IV}$ Also at: Dipartimento DET del Politecnico di Torino, Turin, Italy\\
$^{\rm V}$ Also at: Department of Applied Physics, Aligarh Muslim University, Aligarh, India\\
$^{\rm VI}$ Also at: Institute of Theoretical Physics, University of Wroclaw, Poland\\
$^{\rm VII}$ Also at: An institution covered by a cooperation agreement with CERN\\

\section*{Collaboration Institutes}

$^{1}$ A.I. Alikhanyan National Science Laboratory (Yerevan Physics Institute) Foundation, Yerevan, Armenia\\
$^{2}$ AGH University of Science and Technology, Cracow, Poland\\
$^{3}$ Bogolyubov Institute for Theoretical Physics, National Academy of Sciences of Ukraine, Kiev, Ukraine\\
$^{4}$ Bose Institute, Department of Physics  and Centre for Astroparticle Physics and Space Science (CAPSS), Kolkata, India\\
$^{5}$ California Polytechnic State University, San Luis Obispo, California, United States\\
$^{6}$ Central China Normal University, Wuhan, China\\
$^{7}$ Centro de Aplicaciones Tecnol\'{o}gicas y Desarrollo Nuclear (CEADEN), Havana, Cuba\\
$^{8}$ Centro de Investigaci\'{o}n y de Estudios Avanzados (CINVESTAV), Mexico City and M\'{e}rida, Mexico\\
$^{9}$ Chicago State University, Chicago, Illinois, United States\\
$^{10}$ China Institute of Atomic Energy, Beijing, China\\
$^{11}$ Chungbuk National University, Cheongju, Republic of Korea\\
$^{12}$ Comenius University Bratislava, Faculty of Mathematics, Physics and Informatics, Bratislava, Slovak Republic\\
$^{13}$ COMSATS University Islamabad, Islamabad, Pakistan\\
$^{14}$ Creighton University, Omaha, Nebraska, United States\\
$^{15}$ Department of Physics, Aligarh Muslim University, Aligarh, India\\
$^{16}$ Department of Physics, Pusan National University, Pusan, Republic of Korea\\
$^{17}$ Department of Physics, Sejong University, Seoul, Republic of Korea\\
$^{18}$ Department of Physics, University of California, Berkeley, California, United States\\
$^{19}$ Department of Physics, University of Oslo, Oslo, Norway\\
$^{20}$ Department of Physics and Technology, University of Bergen, Bergen, Norway\\
$^{21}$ Dipartimento di Fisica, Universit\`{a} di Pavia, Pavia, Italy\\
$^{22}$ Dipartimento di Fisica dell'Universit\`{a} and Sezione INFN, Cagliari, Italy\\
$^{23}$ Dipartimento di Fisica dell'Universit\`{a} and Sezione INFN, Trieste, Italy\\
$^{24}$ Dipartimento di Fisica dell'Universit\`{a} and Sezione INFN, Turin, Italy\\
$^{25}$ Dipartimento di Fisica e Astronomia dell'Universit\`{a} and Sezione INFN, Bologna, Italy\\
$^{26}$ Dipartimento di Fisica e Astronomia dell'Universit\`{a} and Sezione INFN, Catania, Italy\\
$^{27}$ Dipartimento di Fisica e Astronomia dell'Universit\`{a} and Sezione INFN, Padova, Italy\\
$^{28}$ Dipartimento di Fisica `E.R.~Caianiello' dell'Universit\`{a} and Gruppo Collegato INFN, Salerno, Italy\\
$^{29}$ Dipartimento DISAT del Politecnico and Sezione INFN, Turin, Italy\\
$^{30}$ Dipartimento di Scienze MIFT, Universit\`{a} di Messina, Messina, Italy\\
$^{31}$ Dipartimento Interateneo di Fisica `M.~Merlin' and Sezione INFN, Bari, Italy\\
$^{32}$ European Organization for Nuclear Research (CERN), Geneva, Switzerland\\
$^{33}$ Faculty of Electrical Engineering, Mechanical Engineering and Naval Architecture, University of Split, Split, Croatia\\
$^{34}$ Faculty of Engineering and Science, Western Norway University of Applied Sciences, Bergen, Norway\\
$^{35}$ Faculty of Nuclear Sciences and Physical Engineering, Czech Technical University in Prague, Prague, Czech Republic\\
$^{36}$ Faculty of Physics, Sofia University, Sofia, Bulgaria\\
$^{37}$ Faculty of Science, P.J.~\v{S}af\'{a}rik University, Ko\v{s}ice, Slovak Republic\\
$^{38}$ Frankfurt Institute for Advanced Studies, Johann Wolfgang Goethe-Universit\"{a}t Frankfurt, Frankfurt, Germany\\
$^{39}$ Fudan University, Shanghai, China\\
$^{40}$ Gangneung-Wonju National University, Gangneung, Republic of Korea\\
$^{41}$ Gauhati University, Department of Physics, Guwahati, India\\
$^{42}$ Helmholtz-Institut f\"{u}r Strahlen- und Kernphysik, Rheinische Friedrich-Wilhelms-Universit\"{a}t Bonn, Bonn, Germany\\
$^{43}$ Helsinki Institute of Physics (HIP), Helsinki, Finland\\
$^{44}$ High Energy Physics Group,  Universidad Aut\'{o}noma de Puebla, Puebla, Mexico\\
$^{45}$ Horia Hulubei National Institute of Physics and Nuclear Engineering, Bucharest, Romania\\
$^{46}$ Indian Institute of Technology Bombay (IIT), Mumbai, India\\
$^{47}$ Indian Institute of Technology Indore, Indore, India\\
$^{48}$ INFN, Laboratori Nazionali di Frascati, Frascati, Italy\\
$^{49}$ INFN, Sezione di Bari, Bari, Italy\\
$^{50}$ INFN, Sezione di Bologna, Bologna, Italy\\
$^{51}$ INFN, Sezione di Cagliari, Cagliari, Italy\\
$^{52}$ INFN, Sezione di Catania, Catania, Italy\\
$^{53}$ INFN, Sezione di Padova, Padova, Italy\\
$^{54}$ INFN, Sezione di Pavia, Pavia, Italy\\
$^{55}$ INFN, Sezione di Torino, Turin, Italy\\
$^{56}$ INFN, Sezione di Trieste, Trieste, Italy\\
$^{57}$ Inha University, Incheon, Republic of Korea\\
$^{58}$ Institute for Gravitational and Subatomic Physics (GRASP), Utrecht University/Nikhef, Utrecht, Netherlands\\
$^{59}$ Institute of Experimental Physics, Slovak Academy of Sciences, Ko\v{s}ice, Slovak Republic\\
$^{60}$ Institute of Physics, Homi Bhabha National Institute, Bhubaneswar, India\\
$^{61}$ Institute of Physics of the Czech Academy of Sciences, Prague, Czech Republic\\
$^{62}$ Institute of Space Science (ISS), Bucharest, Romania\\
$^{63}$ Institut f\"{u}r Kernphysik, Johann Wolfgang Goethe-Universit\"{a}t Frankfurt, Frankfurt, Germany\\
$^{64}$ Instituto de Ciencias Nucleares, Universidad Nacional Aut\'{o}noma de M\'{e}xico, Mexico City, Mexico\\
$^{65}$ Instituto de F\'{i}sica, Universidade Federal do Rio Grande do Sul (UFRGS), Porto Alegre, Brazil\\
$^{66}$ Instituto de F\'{\i}sica, Universidad Nacional Aut\'{o}noma de M\'{e}xico, Mexico City, Mexico\\
$^{67}$ iThemba LABS, National Research Foundation, Somerset West, South Africa\\
$^{68}$ Jeonbuk National University, Jeonju, Republic of Korea\\
$^{69}$ Johann-Wolfgang-Goethe Universit\"{a}t Frankfurt Institut f\"{u}r Informatik, Fachbereich Informatik und Mathematik, Frankfurt, Germany\\
$^{70}$ Korea Institute of Science and Technology Information, Daejeon, Republic of Korea\\
$^{71}$ KTO Karatay University, Konya, Turkey\\
$^{72}$ Laboratoire de Physique des 2 Infinis, Ir\`{e}ne Joliot-Curie, Orsay, France\\
$^{73}$ Laboratoire de Physique Subatomique et de Cosmologie, Universit\'{e} Grenoble-Alpes, CNRS-IN2P3, Grenoble, France\\
$^{74}$ Lawrence Berkeley National Laboratory, Berkeley, California, United States\\
$^{75}$ Lund University Department of Physics, Division of Particle Physics, Lund, Sweden\\
$^{76}$ Nagasaki Institute of Applied Science, Nagasaki, Japan\\
$^{77}$ Nara Women{'}s University (NWU), Nara, Japan\\
$^{78}$ National and Kapodistrian University of Athens, School of Science, Department of Physics , Athens, Greece\\
$^{79}$ National Centre for Nuclear Research, Warsaw, Poland\\
$^{80}$ National Institute of Science Education and Research, Homi Bhabha National Institute, Jatni, India\\
$^{81}$ National Nuclear Research Center, Baku, Azerbaijan\\
$^{82}$ National Research and Innovation Agency - BRIN, Jakarta, Indonesia\\
$^{83}$ Niels Bohr Institute, University of Copenhagen, Copenhagen, Denmark\\
$^{84}$ Nikhef, National institute for subatomic physics, Amsterdam, Netherlands\\
$^{85}$ Nuclear Physics Group, STFC Daresbury Laboratory, Daresbury, United Kingdom\\
$^{86}$ Nuclear Physics Institute of the Czech Academy of Sciences, Husinec-\v{R}e\v{z}, Czech Republic\\
$^{87}$ Oak Ridge National Laboratory, Oak Ridge, Tennessee, United States\\
$^{88}$ Ohio State University, Columbus, Ohio, United States\\
$^{89}$ Physics department, Faculty of science, University of Zagreb, Zagreb, Croatia\\
$^{90}$ Physics Department, Panjab University, Chandigarh, India\\
$^{91}$ Physics Department, University of Jammu, Jammu, India\\
$^{92}$ Physics Department, University of Rajasthan, Jaipur, India\\
$^{93}$ Physics Program and International Institute for Sustainability with Knotted Chiral Meta Matter (SKCM2), Hiroshima University, Hiroshima, Japan\\
$^{94}$ Physikalisches Institut, Eberhard-Karls-Universit\"{a}t T\"{u}bingen, T\"{u}bingen, Germany\\
$^{95}$ Physikalisches Institut, Ruprecht-Karls-Universit\"{a}t Heidelberg, Heidelberg, Germany\\
$^{96}$ Physik Department, Technische Universit\"{a}t M\"{u}nchen, Munich, Germany\\
$^{97}$ Politecnico di Bari and Sezione INFN, Bari, Italy\\
$^{98}$ Research Division and ExtreMe Matter Institute EMMI, GSI Helmholtzzentrum f\"ur Schwerionenforschung GmbH, Darmstadt, Germany\\
$^{99}$ Saha Institute of Nuclear Physics, Homi Bhabha National Institute, Kolkata, India\\
$^{100}$ School of Physics and Astronomy, University of Birmingham, Birmingham, United Kingdom\\
$^{101}$ Secci\'{o}n F\'{\i}sica, Departamento de Ciencias, Pontificia Universidad Cat\'{o}lica del Per\'{u}, Lima, Peru\\
$^{102}$ Stefan Meyer Institut f\"{u}r Subatomare Physik (SMI), Vienna, Austria\\
$^{103}$ SUBATECH, IMT Atlantique, Nantes Universit\'{e}, CNRS-IN2P3, Nantes, France\\
$^{104}$ Suranaree University of Technology, Nakhon Ratchasima, Thailand\\
$^{105}$ Technical University of Ko\v{s}ice, Ko\v{s}ice, Slovak Republic\\
$^{106}$ The Henryk Niewodniczanski Institute of Nuclear Physics, Polish Academy of Sciences, Cracow, Poland\\
$^{107}$ The University of Texas at Austin, Austin, Texas, United States\\
$^{108}$ Universidad Aut\'{o}noma de Sinaloa, Culiac\'{a}n, Mexico\\
$^{109}$ Universidade de S\~{a}o Paulo (USP), S\~{a}o Paulo, Brazil\\
$^{110}$ Universidade Estadual de Campinas (UNICAMP), Campinas, Brazil\\
$^{111}$ Universidade Federal do ABC, Santo Andre, Brazil\\
$^{112}$ University of Cape Town, Cape Town, South Africa\\
$^{113}$ University of Houston, Houston, Texas, United States\\
$^{114}$ University of Jyv\"{a}skyl\"{a}, Jyv\"{a}skyl\"{a}, Finland\\
$^{115}$ University of Kansas, Lawrence, Kansas, United States\\
$^{116}$ University of Liverpool, Liverpool, United Kingdom\\
$^{117}$ University of Science and Technology of China, Hefei, China\\
$^{118}$ University of South-Eastern Norway, Kongsberg, Norway\\
$^{119}$ University of Tennessee, Knoxville, Tennessee, United States\\
$^{120}$ University of the Witwatersrand, Johannesburg, South Africa\\
$^{121}$ University of Tokyo, Tokyo, Japan\\
$^{122}$ University of Tsukuba, Tsukuba, Japan\\
$^{123}$ University Politehnica of Bucharest, Bucharest, Romania\\
$^{124}$ Universit\'{e} Clermont Auvergne, CNRS/IN2P3, LPC, Clermont-Ferrand, France\\
$^{125}$ Universit\'{e} de Lyon, CNRS/IN2P3, Institut de Physique des 2 Infinis de Lyon, Lyon, France\\
$^{126}$ Universit\'{e} de Strasbourg, CNRS, IPHC UMR 7178, F-67000 Strasbourg, France, Strasbourg, France\\
$^{127}$ Universit\'{e} Paris-Saclay Centre d'Etudes de Saclay (CEA), IRFU, D\'{e}partment de Physique Nucl\'{e}aire (DPhN), Saclay, France\\
$^{128}$ Universit\`{a} degli Studi di Foggia, Foggia, Italy\\
$^{129}$ Universit\`{a} del Piemonte Orientale, Vercelli, Italy\\
$^{130}$ Universit\`{a} di Brescia, Brescia, Italy\\
$^{131}$ Variable Energy Cyclotron Centre, Homi Bhabha National Institute, Kolkata, India\\
$^{132}$ Warsaw University of Technology, Warsaw, Poland\\
$^{133}$ Wayne State University, Detroit, Michigan, United States\\
$^{134}$ Westf\"{a}lische Wilhelms-Universit\"{a}t M\"{u}nster, Institut f\"{u}r Kernphysik, M\"{u}nster, Germany\\
$^{135}$ Wigner Research Centre for Physics, Budapest, Hungary\\
$^{136}$ Yale University, New Haven, Connecticut, United States\\
$^{137}$ Yonsei University, Seoul, Republic of Korea\\
$^{138}$  Zentrum  f\"{u}r Technologie und Transfer (ZTT), Worms, Germany\\
$^{139}$ Affiliated with an institute covered by a cooperation agreement with CERN\\
$^{140}$ Affiliated with an international laboratory covered by a cooperation agreement with CERN.\\

\end{flushleft} 

\end{document}